\documentclass[11pt]{article}
\usepackage{amssymb,amsthm,amsmath,amsfonts}
\usepackage{epsfig}

\theoremstyle{plain}
\begingroup
 
\newtheorem{theorem}{Theorem} 
\newtheorem{corollary}{Corollary}
\newtheorem{lemma}{Lemma}
\endgroup

\theoremstyle{definition}
\newtheorem{remark}{Remark}

\newcommand{\nwc}{\newcommand}
\nwc{\bit}{\begin{itemize}}
\nwc{\eit}{\end{itemize}}
\nwc{\Levy}{L\'evy}
\nwc{\LK}{L\'evy-Khintchine}
\nwc{\LI}{L\'evy-It\^{o}}
\nwc{\CH}{Cole-Hopf}
\nwc{\Holder}{H\"{o}lder}
\nwc{\cadlag}{c\`{a}dl\`{a}g}
\nwc{\be}{\begin{equation}}
\nwc{\ee}{\end{equation}}
\nwc{\ba}{\begin{eqnarray}}
\nwc{\ea}{\end{eqnarray}}
\nwc{\la}{\label}
\nwc{\nn}{\nonumber}
\nwc{\Z}{\mathbb{Z}}
\nwc{\C}{\mathbb{C}}
\nwc{\E}{\mathbb{E}}
\nwc{\R}{\mathbb{R}}
\nwc{\N}{\mathbb{N}}
\nwc{\prob}{\mathbb{P}}
\nwc{\Skor}{\mathbb{D}}
\nwc{\attr}{\mathcal{A}}
\nwc{\PP}{\mathcal{P}}
\nwc{\PPE}{\mathcal{P}(E)}
\nwc{\M}{\mathcal{M}}
\nwc{\Tt}{T^{(t)}}
\nwc{\Ut}{U^{(t)}}
\nwc{\Vt}{V^{(t)}}
\nwc{\tlambda}{\tilde{\lambda}}
\nwc{\tbar}{\bar{t}}
\nwc{\gtx}{g^{(t)}_x}
\nwc{\order}{\prec}
\nwc{\law}{\stackrel{\mathcal{L}}{\rightarrow}}
\nwc{\eqd}{\stackrel{\mathcal{L}}{=}}
\nwc{\vp}{\varphi}
\nwc{\Vp}{\Phi}
\nwc{\psilevy}{\Psi}
\nwc{\ve}{\varepsilon}
\nwc{\veps}{\varepsilon}
\nwc{\eps}{\ve}
\nwc{\betarsc}{\theta}
\nwc{\cl}{c\'{a}dl\'{a}g}
\nwc{\qref}[1]{(\ref{#1})}
\nwc{\D}{\partial}
\nwc{\Ebar}{\bar{E}}
\nwc{\mmt}{m}
\nwc{\dnto}{\downarrow}
\nwc{\fzero}{F_\rho} 
\nwc{\fone}{M_\rho} 
\nwc{\ip}[1]{\langle #1 \rangle}
\nwc{\ipbig}[1]{\left\langle #1 \right\rangle}
\nwc{\Lip}{\mathop{\rm Lip}\nolimits}
\nwc{\Tmin}{T_{\min}}
\nwc{\Tmax}{T_{\max}}
\nwc{\Tgel}{T_{\rm gel}}
\nwc{\LL}{\mathcal{L}}
\nwc{\mudot}{\mu}
\nwc{\nudot}{\nu}
\nwc{\rme}{{\rm e}}
\nwc{\rmi}{{\rm i}}
\nwc{\nsup}{^{(n)}}
\nwc{\ksup}{^{(k)}}
\nwc{\jsup}{^{(j)}}
\nwc{\nksup}{^{(n_k)}}
\nwc{\inv}{^{-1}}
\nwc{\qxfac}{(1-\rme^{-qx})}
\nwc{\one}{\mathbf{1}}
\nwc{\psib}{\psi^{(b)}}
\nwc{\pss}{\psi_\#}
\nwc{\phss}{\Phi_\#}
\nwc{\asharp}{\gamma}
\nwc{\vfour}{\hat{v}}
\nwc{\Pss}{\Psi_\#}
\nwc{\pibar}{\Pi}
\nwc{\Ltail}{\bar{\Lambda}}
\nwc{\vb}{v^{(b)}}
\nwc{\spec}{\tilde{\beta}}
\nwc{\hv}{\hat{v}}
\nwc{\hvl}{\hat{v}_L}
\nwc{\lap}{{\,\cal L}}

\begin{document}
\title{Universality classes in Burgers turbulence} 
\author{Govind Menon\textsuperscript{1} and Robert. L.
Pego\textsuperscript{2}}

\date{\today}

\maketitle
\begin{abstract}
We establish necessary and sufficient conditions for the shock statistics
to approach self-similar form in Burgers turbulence with \Levy\/
process initial data. The proof relies upon an elegant closure theorem
of Bertoin 
and Carraro and Duchon 
that reduces the
study of shock statistics to  Smoluchowski's coagulation equation with
additive kernel, and upon our previous characterization of the 
domains of attraction
of self-similar solutions for this equation.  
\end{abstract}

  \Levy\/ processes, dynamic scaling, regular variation, 
agglomeration, coagulation, coalescence, shock statistics.

\medskip
\footnotetext[1]
{Division of Applied Mathematics, Box F, Brown University, Providence, RI 02912.
Email: menon@dam.brown.edu}
\footnotetext[2]{Center for Nonlinear Analysis, Department of
  Mathematical Sciences,  Carnegie Mellon University, Pittsburgh, PA 15213.
Email: rpego@cmu.edu}

\pagebreak
\section{Introduction}
The construction of stochastic processes that are also weak solutions
to the equations of fluid mechanics is one approach to rigorous
mathematical theories of turbulence. This is poorly understood at
present, and we must 
settle for insights from vastly simplified model problems. We consider
the invisicid Burgers equation  
\begin{equation}
\label{E.burgers}
\D_t u + \D_x\left( \frac{u^2}{2} \right) =0, \quad t>0,\ x \in \R, \quad
u(x,0)=u_0(x),
\end{equation}
with random initial data $u_0$. The problem is to determine the
statistical properties of the \CH\ (entropy) solution $u(x,t)$ to
(\ref{E.burgers}), given the statistical properties of $u_0$. 
There is a large literature on the subject; we refer to
Burgers' book~\cite{Burgers} and the
more recent survey articles~\cite{ES,Gurbatov2,Woc}. 
The problem was proposed by Burgers as a
model for turbulence in incompressible fluids, but it has several
well-known flaws in this regard. 

Explicit solutions play a special role in the theory.
Burgers studied the case when $u_0$ is 
white noise in his monograph~\cite{Burgers}.
His work remains the foundation for several rigorous results, which
culminate with the 
complete solution by Frachebourg and Martin for the velocity and shock
statistics (see ~\cite{Frachebourg} and references therein).
The case when $u_0$ is a Brownian motion has attracted much attention
since the work of  
She, Aurell and Frisch~\cite{She} and Sinai~\cite{Sinai}. 
An elegant solution to
this problem was obtained by Bertoin~\cite{B_burgers} and Carraro and
Duchon~\cite{Duchon1}. More generally, these authors considered 
initial data that comprise a \Levy\/ process
with only downward jumps (i.e., shocks). A \Levy\/ process  
$X_x$ ($x \geq 0$) is a continuous-time random walk with stationary and
independent increments. It is determined completely 
by its characteristic exponent $\Psi$, satisfying $\E(e^{ikX_x})
= e^{-x\Psi(k)}$, via the celebrated \LK\/ formula 
\be 
\label{E.LK1}
\Psi(k)  = i b k +
\frac{\sigma^2 k^2}{2} + \int_{\R} \left(1-e^{iks}+ 
i ks\one_{|s|<1} \right) \Pi(ds), \quad k \in \R.
\ee
The process $X$ is the superposition of three independent
processes related to this formula: a Brownian motion
with variance $\sigma^2$ and drift $b \in \R$, a compound Poisson
process with jump measure $\Pi \one_{|s| \geq 1}$, and a pure 
jump martingale 
with jump measure $\Pi \one_{|s| <1}$ (see~\cite[Ch.1]{B_book}).
The measure $\Pi$ is arbitrary, subject to the condition
$\int_\R (1\wedge s^2)\Pi(ds)<\infty$, where $a\wedge b$ means $\min(a,b)$.

We say $X$ is {\em spectrally negative\/} if $\Pi$ is
concentrated on the half-line $s<0$. In all that follows we assume
\be
\la{E.initial}
u_0(x)  = \left\{ \begin{array}{l} 0, \quad x < 0, \\
\text{a spectrally negative \Levy\/ process}, \quad x
\geq  0, \end{array} \right.  
\ee
As we show below, we can always reduce to the case where $u_0(x)$ has zero mean
$\E(u_0(x))$ for all $x$, 
and $\int_{-\infty}^0(|s|\wedge s^2) \Pi(ds) < \infty$. It is then 
more convenient to use the Laplace exponent 
\be
\label{E.LKu}
 \psi(q) = -\Psi(-iq) = \frac{\sigma^2 q^2}{2}  + \int_0^\infty \left(
e^{-qs} -1 + qs \right) \,\Lambda(ds), \quad q >0,
\ee
where  $\Lambda((s,\infty)) = \Pi((-\infty,-s))$ for every $s>0$, so that 
$\E(e^{qX_x}) = e^{x\psi(q)}$ and $\E(X_x)=x\psi'(0)=0$.

For this class of initial data, Bertoin
proved a remarkable closure property for the entropy
solution of \qref{E.burgers}, namely: 
$x\mapsto u(x,t)-u(0,t)$  {\em remains a spectrally negative
\Levy\/ process for all $t>0$.}  
This closure property was first noted by Carraro and
Duchon in connection with their notion of statistical
solutions to Burgers equation~\cite{Duchon0}. That these
statistical solutions agree with the \CH\/ solution for spectrally
negative data  was shown by Bertoin~\cite[Thm. 2]{B_burgers}.  
The closure property fails if $u_0$ has positive jumps---These
positive jumps open into rarefaction waves for $t>0$, and this is
incompatible with the rigidity of sample paths of \Levy\/ processes.
An interesting formal analysis of closure properties of Burgers equation
is presented in~\cite{Duchon2}. 

Henceforth, we write $v(x,t)=u(x,t)-u(0,t)$ for brevity.  
The \LK\ representation now implies that the law of the \Levy\/ process
$x\mapsto v(x,t)$ is completely described by a corresponding 
``\Levy\/ triplet'' $(b_t,\sigma_t^2, \Lambda_t)$. 
The mean drift $b_t=\E(v(1,t))$ satisfies $b_t=b_0=0$ for every 
$t \geq 0$. Moreover, for every $t >0$, $v(\cdot,t)$ is
of bounded variation, thus the variance $\sigma_t=0$.  Consequently, 
the law of $v(\cdot,t)$ is completely determined
by only the jump measure $\Lambda_t$ which contains the shock statistics.

It is a striking fact, implicit in \cite{B_burgers}, that the evolution of
$\Lambda_t$ is described by {\em Smoluchowski's coagulation equation with
additive kernel}, an equation that arises in entirely different
areas such as the analysis of algorithms~\cite{Chassaing}, the kinetics of
polymerization~\cite{Ziff}, and cloud formation from
droplets~\cite{Golovin} (see~\cite{Aldous} for a review). 
What this means is that {\em mean-field theory is exact\/} for Burgers
equation with   
initial data of the form \qref{E.initial}, i.e., random one-sided
data with stationary and independent increments.
We give a precise statement to this effect below in 
Theorem~\ref{T.meanfield}. Several connections
between stochastic models of coalescence and Burgers turbulence are
reviewed in~\cite{B_icm}. 

Here, we use the closure property as a basis for a rigorous study of
{\em universality classes for dynamic scaling} in Burgers turbulence. 
Our motivation is the 
following. A central theme in studies of homogeneous isotropic turbulence in
incompressible fluids is the universality of the Kolmogorov 
spectrum~\cite{Kolm}. A possible rigorous formulation of such 
universality involves (a) the construction of stochastic
processes that mimic a `typical turbulent flow', and (b) a characterization
of the domains of attraction of these processes.  
For Burgers turbulence,
step (a) consists of constructing exact solutions for special initial data, say
white noise or Brownian motion. In this article, we
carry out step (b) for initial data that satisfy \qref{E.initial}. 

Domains of attraction are studied in the classical limit theorems in  
probability (e.g., the central limit theorem), and their process
versions (e.g., Donsker's invariance principle). For Smoluchowski's
coagulation equation with additive kernel, we characterized 
all possible domains of attraction in~\cite{MP1}, 
a result akin to the classical limit theorems. 
In this article we deal with a process version. 
In all that follows, we consider the processes
$x\mapsto v(x,t)$ as elements
of the space $\Skor$ of right continuous paths $\R_+\to \R$ 
with left limits (\cadlag\/ paths) equipped with the Skorokhod
topology~\cite[Ch. VI]{Jacod}. The shock statistics determine
completely the law of this process (a probability measure on
$\Skor$). Approach to limiting forms will be phrased in terms of 
weak convergence of probability measures on $\Skor$. 

Among the initial data we consider, the stable processes are of particular
importance because of their self-similarity. Let $X^\alpha, \alpha \in
(1,2]$ denote the stable process with Laplace exponent $q^\alpha$ 
($\alpha=2$ corresponds to Brownian motion). 
The corresponding jump measure 
$\Lambda(ds)=s^{-1-\alpha}\,ds/\Gamma(-\alpha)$ for $\alpha<2$.
There is a one-to-one correspondence between (a) these stable 
processes, (b) statistically self-similar solutions in
Burgers turbulence, and (c) self-similar solutions  to Smoluchowski's
coagulation equation. 
Precisely, this works as follows. Let $\alpha \in (1,2]$, and let $T^\alpha$
denote the first-passage process for $x\mapsto X^\alpha_x+x$, i.e.,
\be
\label{E.firstpass}
T^\alpha_x = \inf \{y \left| X^\alpha_y + y > x \right.\}.
\ee
Then for the solution to (\ref{E.burgers}) with 
$u_0(x)=X^\alpha_x$ for $x \geq 0$, 
$v(x,t)$ is statistically self-similar, with 
\begin{equation}
\label{eq:selfsimilar}
v(x,t) \eqd 
t^{1/\beta-1} V^\alpha_{xt^{-1/\beta}}
:= 
t^{1/\beta-1}\left(xt^{-1/\beta} -
T^\alpha_{xt^{-1/\beta}} \right) 
, \quad \; t,x>0,
\end{equation}
where $\beta= (\alpha-1)/\alpha$.
Here $\eqd$ means both processes define the same measure on $\Skor$.
The process $T^\alpha$ is a pure jump \Levy\ process with
\Levy\ measure $f_\alpha(s) 
\,ds$, where $f_\alpha$ is the number density profile
of a self-similar solution to Smoluchowski's coagulation
equation~\cite[Sec 6]{MP1}:
\be
\label{E.ndensity}
f_\alpha(s) = \frac{1}{\pi} \sum_{k=1}^\infty \frac{(-1)^{k-1}
  s^{k\beta-2}}{k!} \Gamma(1+k-k \beta) \sin \pi k \beta, \quad \alpha
\in (1,2].
\ee
These solutions are related to classical distributions in probability 
theory by rescaling. If  $p(s;\alpha,2-\alpha)$ denotes the density of
a  maximally-skewed \Levy\ stable  law~\cite[XVII.7]{Feller} we have
\cite{B_eternal,MP1} 
\be
\label{E.max_stable}
f_\alpha (s)  = s^{\beta-2} p(s^\beta;\alpha, 2-\alpha).
\ee
By the \LI\ decomposition~\cite[Thm 1.1]{B_book} and
(\ref{eq:selfsimilar}) we may conclude   
that the magnitudes of shocks  in $u(\cdot,t)$ form a Poisson point
process valued in $(0,\infty)$ whose characteristic measure is
\begin{equation}
\label{eq:shock_intensity}
\Lambda^\alpha_t(ds) = t^{1-2/\beta} 
f_\alpha\left( {s t^{1-1/\beta}} \right) \, ds.
\end{equation}

For $1<\alpha <2$ the self-similar solutions have algebraic tails,
with $f_\alpha(s) \sim s^{-1-\alpha}/\Gamma(-\alpha)$ as $s\to\infty$.
The case $\alpha=2$ is particularly important since it corresponds to Brownian
initial data. Here we obtain a solution found by Golovin in a model
for cloud formation from droplets~\cite{Golovin}, 
\be
\label{E.golovin}
f_2(s) = (4\pi)^{-1/2} s^{-3/2} e^{-s/4}.
\ee
For the corresponding solution to \qref{E.burgers}, the
law of $v(x,t)$ can be recovered from 
the law of $T^2_x$, the first-passage time for Brownian motion with
unit drift, which is explicitly given as follows (see section 2.6):
\be
\label{E.golovin2}
{\prob}(T^2_x \in (y,y+dy) ) =
\frac{x  \one_{y >0}}{2\sqrt{\pi y^3}}
\exp\left(-\frac{(x-y)^2}{4y}\right) \, dy. 
\ee


Considering now arbitrary solutions to \qref{E.burgers} with initial data 
\qref{E.initial}, we classify solutions that approach self-similar
form as $t\to\infty$ as follows.  We establish necessary and
sufficient conditions for convergence of the laws of the rescaled processes 
\begin{equation}
\label{E.rscV}
x\mapsto \Vt_x =  \frac{t}{\lambda(t)}v(\lambda(t)x, t)
\end{equation}
in the sense of weak convergence of measures on $\Skor$.  (Since
the shocks coalesce, a rescaling $\lambda(t) \to \infty$ is needed
to obtain a non-trivial limit. The amplitude scaling is natural, 
see Section~\ref{S.proof}.)  Convergence
to a process $V^*$ is written $\Vt \law V^*$ as in~\cite{Jacod}.
Recall that a positive function $L$ is said to be slowly varying at $\infty$
if $\lim_{t\to\infty}L(tx)/L(t)=1$ for all $x>0$. 
\begin{theorem}
\label{T.main}
Let $u_0$ be a spectrally negative \Levy\ process with zero mean 
$\E(u_0(x))$, 
variance $\sigma_0^2\ge0$, and downward jump measure 
satisfying $\int_0^\infty (s \wedge s^2) \Lambda_0(ds)<\infty$.
\begin{enumerate}
\item Suppose there is a rescaling  $\lambda(t) \rightarrow \infty$ as
  $t\to\infty$ and a \Levy\ process $V^*$ with zero mean $\E(V^*_1)$ 
such that the random variables $\Vt_1$ converge to $V^*_1$ in law.  
Then there exists $\alpha \in (1,2]$ 
and a function $L$ slowly varying at infinity such that 
\begin{equation}
\label{E.RV}
\sigma_0^2+ \int_0^s r^2 \Lambda_0(dr) \sim {s^{2-\alpha} L(s)}
\quad\text{as $ s \to \infty.$} 
\end{equation}
\item Conversely, assume that there exists $\alpha \in (1,2]$ and a
function $L$ slowly varying at infinity such that (\ref{E.RV})
  holds. Then there is a strictly increasing rescaling $\lambda(t) \to
  \infty$ such that $\Vt \law V^{\alpha}$.
\end{enumerate}
\end{theorem}
\begin{remark}
Since $\Vt$ and $V^*$ are \Levy\/ processes, we have 
$\Vt \law V^*$ if and only if
we have convergence in law of the random variables
$\Vt_{x_0}$ for some fixed $x_0 \in (0,\infty)$ (see \qref{E.eqv1}-\qref{E.eqv2} in
section 3 below).  We take $x_0=1$ without loss of generality. 
Part (2) implies in particular that the only
possible limits are statistically self-similar.
\end{remark}
\begin{remark}
\label{R.energy}
We say a solution has finite energy if for any finite
interval $I \subset \R_+$ we have $\E\left(\int_I|v(x,t)|^2 \, dx\right) <
\infty$. The jump measure $\Lambda_t$ for the solution is related 
to the energy by (see Section~\ref{S.energy})
\[ \E\left(\int_I v(x,t)^2 \, dx \right) = \left(\int_0^\infty s^2
\Lambda_t(ds) \right) \int_I x\,dx. \]
The integral in \qref{E.RV} is thus a measure of the energy in an
interval. If it is initially finite, it is conserved for $t>0$,
and it remains infinite if it is initially infinite. 
The only self-similar solution  with finite energy corresponds to
$\alpha=2$, and Theorem~\ref{T.main} implies it attracts all solutions with
initially finite energy. In this sense, one may say that the finite energy
solution is universal. However, Theorem~\ref{T.main} also indicates 
the delicate dependence of the domains of attraction on the tail
behavior of $\Lambda_0$. Heavy-tailed solutions seem to us no less interesting
than those with finite energy.  Finer results on asymptotics, and a
compactness theorem for subsequential limits that 
builds on Bertoin's \LK\/ classification for eternal
solutions to Smoluchowski's equation~\cite{B_eternal}, will be
developed elsewhere. 
\end{remark}
\begin{remark}
The case of zero mean, $b_0=0$, is the most interesting. 
If $b_0> 0$ or $b_0<0$ we can
reduce to this case by a change of variables (see below). If $b_0<0$,
the solution is defined only for $0 \leq t <
-b_0^{-1}$. Theorem~\ref{T.main} then characterizes the approach to
self-similarity at the blow-up time. If $b_0>0$ then the behavior of
the solution as $t\to\infty$ is determined by the zero-mean solution
with the same $\sigma_0^2$ and $\Lambda_0$ at the finite time $b_0^{-1}$.
\end{remark}
\begin{remark} The \CH\/ solution is  geometric  and 
Theorem~\ref{T.main} may be a viewed as a limit theorem for
statistics of minima. The utility of regular variation in such
problems is widely known~\cite{Resnick}. If the initial data is white
noise, the \CH\/ solution is a study of the parabolic hull of
Brownian motion. Groeneboom's work on this problem~\cite{Groen} is the
basis for several results on Burgers turbulence~(in
particular~\cite{AE,Frachebourg,Giraud}).  We have been
unable to find a   similar reference to the problem we consider in the
probability literature (\cite{Bingham2} seems the closest). 
\end{remark}

\begin{remark}
There is a growing literature on  intermittence, and the asymptotic
self-similarity of Burgers turbulence, see for
example~\cite{Woc2,Gurbatov}.  
Numerical simulations and heuristic arguments suggest that this is a
subtle problem with several distinct regimes. 
It is hard to obtain rigorous results for general initial data. 
Theorem~\ref{T.main} tells us that the approach
to self-similarity is at least as complex as in the classical limit
theorems of probability.  
\end{remark}

The rest of this article is organized as follows. We explain the
mapping from Burgers equation to Smoluchowski's coagulation
equation in Section~\ref{S.meanfield}. This is followed by the proof of
Theorem~\ref{T.main} in Section~\ref{S.proof}. Finally, in
Section~\ref{S.energy} we compute a number of statistics of
physical interest: energy and dissipation in solutions,  the
Fourier-Laplace spectrum, and the multifractal spectrum.

\section{Mean field theory for Burgers equation}
\label{S.meanfield}
In this section we explain the connection between Burgers equation with
spectrally negative \Levy\ process data and Smoluchowski's coagulation
equation. The  main results are due to
Bertoin~\cite{B_burgers} and Carraro and Duchon~\cite{Duchon1}. 
We follow Bertoin's approach, and explain results  implicit
in~\cite{B_burgers} and \cite{B_eternal}. 
We think it worthwhile to make this connection widely known 
in full generality, since the results are of interest to
many non-probabilists. Exact solutions of this simplicity are also
useful as benchmark problems for numerical calculations. 

\subsection{Shock coalescence and Smoluchowski's coagulation equation}
Smoluchowski's coagulation equation is a widely used mean-field model of 
cluster growth~(see~\cite{Aldous,Drake} for introductions).
We begin with a heuristic derivation of
the coagulation equation as a mean-field model of shock
coalescence. First consider the evolution of 
a single shock of size $s>0$.  
Let $u_0(x)= -s \one_{x \geq 0}$. Then the solution is  
\be
\la{E.ba1}
u(x,t) = -s \one_{x \geq x_1(t)}, \quad x_1(t) =
-\frac{s}{2}t. 
\ee
Shock coalescence is nicely seen as follows. 
Let $u_0(x) = -\sum_{k=1}^N s_k(0)
\one_{x \geq x_k(0)}$ where $s_k(0) >0$ for 
$k=1,\ldots,N$ and $x_1(0) < \ldots x_N(0)$. 
The solution may be constructed
using the method of characteristics and the standard jump condition
\[
\dot{x} = \frac12(u^- + u^+)
\]
across a shock at $x=x(t)$, where $u^-$ and $u^+$ denote respectively the left
and right limits of $u(\cdot,t)$ at $x$. At any time
$t>0$, there are $N(t) \leq N(0)$ shocks at locations $x_1(t) <
x_k(t) < x_{N(t)}(t)$ and 
\be
\la{E.ba2}
u(x,t) = - \sum_{k=1}^N s_k(t) \one_{x \geq x_k(t)}, \quad
\dot{x}_k(t_+) = -\sum_{j=1}^{k-1} s_j(t_+) -
\frac{s_k(t_+)}{2}. 
\ee
The shock sizes $s_k(t)$ are constant between collisions, 
and add upon collision---when shocks $k$ and $k+1$ collide,  
we set $s_k(t_+) = s_k(t_-)+ s_{k+1}(t_-)$ and relabel.
This yields an appealing {\em sticky
  particle\/} or {\em ballistic aggregation\/} scenario.
We say a system of particles with position, mass and
velocity $(x_k(t),m_k(t),v_k(t))$ undergoes ballistic aggregation if
(a) the particles move with constant mass and velocity between collisions, 
and (b) at collisions, the colliding particles stick to form a single particle,
conserving mass and momentum in the process. We map this shock coalescence
problem to a sticky particle system by setting $m_k= s_k$ and 
$v_k = \dot x_k$. 
Suppose particles $k$ and $k+1$ meet at time $t$. 
Then, with unprimed variables denoting values before collision and
primed variables denoting values after, 
since $v_{k+1}=v_k-(m_k+m_{k+1})/2$ we have
\[
m_k v_k + m_{k+1} v_{k+1} 
=(m_k + m_{k+1})\left(v_k - \frac{m_{k+1}}{2} \right) 
= m_k'v_k'. 
\]
 
\begin{figure}
\centerline{\epsfysize=5cm{\epsffile{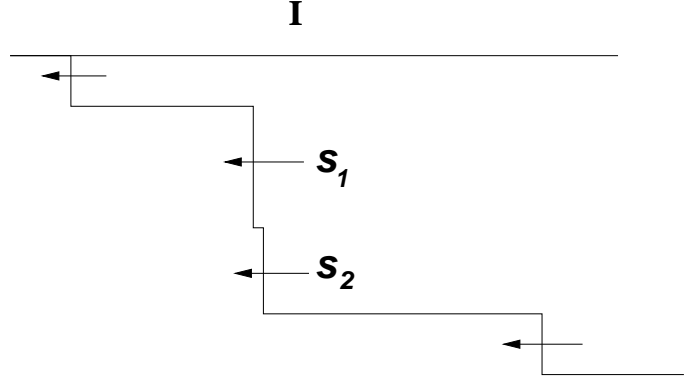}}}
\caption{Binary clustering of shocks \label{fig:heuristic} 
}
\end{figure}

The calculations so far involve no randomness. Suppose now that the 
shock sizes $s_j$ are independent and let $f(s,t)\,ds$ denote 
the expected number of shocks per unit length with size in 
$[s,s+ ds]$.  We derive a mean-field rate equation for $f$ as follows. 
Let $I$ be an interval of unit length. 
The number density changes because of the flux of
shocks entering and leaving $I$ and because of shock collisions within $I$. 
On average, the velocity difference across $I$ is 
\[
M_1(t) = \int_0^\infty s f(s,t) ds,
\]
therefore the average influx is $M_1(t)f(s,t)\,ds$.
Next consider the formation of a shock of size $s_1+ s_2$ by a collision
  of shocks of size $s_1$ and $s_2$ as shown in
  Figure~\ref{fig:heuristic}. The relative velocity 
  between these shocks is $(s_1+ s_2)/2$
  (see Figure~\ref{fig:heuristic}). 
The expected number of neighboring pairs with sizes in
$[s_1,s_1+ds_1]$, 
$[s_2,s_2+ds_2]$ 
respectively is 
\[
f(s_1,t)f(s_2,t) \,ds_1\,ds_2 .
\]
The probability that these neighboring shocks are near enough to collide in
time $dt$ is $\frac12(s_1+s_2)\,dt$, 
thus the number of these shocks 
that collide in time $dt$ is 
\be
\label{E.rate}
f(s_1,t) f(s_2,t) \frac{s_1+ s_2}{2}\, ds_1\,ds_2\,dt. 
\ee
Summing over all collisions that create shocks of size $s=s_1+s_2$,
and accounting for the loss of shocks of size $s$ ($=s_1$ or $s_2$)
in collisions with other shocks, we obtain the rate equation
\[
{\partial_t f (s,t)=  M_1(t) f + Q(f,f)}
\] 
where 
$Q(f,f)$ denotes the collision operator given by  
\[
Q(f,f)(s,t) = \frac{1}{2}\int_0^s s f(s_1,t)
f(s-s_1, t) \, ds_1
 - \int_0^\infty (s+ s_1)  f(s,t) f(s_1,t) \,ds_1. 
\]
We integrate in $s$ to
find $\dot M_1 = M_1^2$, therefore the normalized
density $f/M_1$ satisfies the equation
\be
\la{E.smol_ba}
\frac1{M_1} \partial_t \left( \frac{f}{M_1} \right) = Q \left( \frac{f}{M_1},
\frac{f}{M_1} \right). 
\ee
Up to a change of time scale, this is a fundamental mean-field model of
coalescence: Smoluchowski's coagulation equation with additive kernel. 
We treat this equation in greater depth below. 

More precisely, it turns out that the random solution $u(x,t)$ has the
structure described in \qref{E.ba2} when the initial data $u_0$ consists of a
compound Poisson process with only downward jumps. The mean drift rate at
time $t$ is then $-M_1(t)$, and this example shows that the solution
blows up at the time $M_1(0)^{-1}$ in this case.
We show below (see \qref{E.changedrift}) that one may remove the mean drift by
a change of scale and slope, yielding `sawtooth' data with a deterministic
upward drift that compensates the random downward jumps. 
For such data we obtain a global solution. Thus, there is no essential
distinction between sawtooth data and the decreasing initial data
considered above.




\subsection{The \CH\ formula}
The modern notion of an entropy solution stems from the penetrating
analysis by Hopf of the vanishing viscosity limit to
\qref{E.burgers}. His work was based on a change of variables
(re)discovered independently by Cole 
and Hopf~\cite{Cole,Hopf}.  This solution is obtained via
minimization of the \CH\/ function \cite{Cole,Hopf}
\begin{equation}
\label{E.CH}
H(y,t;x) = \frac{(x-y)^2}{2t} 
+ \int_{-\infty}^y u_0(y')dy'.
\end{equation}
The minimum in $y$ is well-defined for all $t>0$ 
provided $U(y)= \int_0^y u_0(y') dy'$ is
lower semicontinuous and $\lim_{x \to \pm \infty} y^{-2} U(y)=0$.
This is a mild assumption and holds for the random data we consider
provided that the mean drift is zero. 
We denote the extreme points where $H$ is minimized by
\be
\la{E.IL}
a_-(x,t) = \inf\{z | H(z,t;x)= \min_y H\}, \quad 
a_+(x,t) = \sup\{z| H(z,t;x)= \min_y H \}. 
\ee
Notice that any $z \in \R$ such that $x= tu_0(z) +z$ is a critical
point of $H$, and represents a Lagrangian point that arrives at $x$ at
time $t$. Of these $z$, the `correct' Lagrangian points  are 
the minimizers of $H$. If $a_-(x,t)=a_+(x,t)$, this point is unique,
and we have 
 \begin{equation} 
\label{E.hopf}
u(x,t) = \frac{x - a_\pm(x,t)}{t}, \quad x \in \R,\ t >0. \quad
\end{equation}
There is a shock at $(x,t)$ when $a_-(x,t) \neq a_+(x,t)$. In this
case, the Lagrangian interval 
$[a_-(x,t),a_+(x,t)]$ is absorbed into the shock and the velocity of the
shock is given by the Rankine-Hugoniot condition (conservation of momentum)
\be
\la{E.ushock}
u(x,t) =  \frac{u(x_+,t) + u(x_-,t)}{2} = 
\frac{1}{a_+(x,t)-a_-(x,t)}\int_{a_-(x,t)}^{a_+(x,t)} u_0(y) \, dy. 
\ee
It will be convenient for us to assume that $u$ is
right-continuous and we call $a(x,t) = a_+(x,t)$ the {\em inverse
Lagrangian function\/}. Of course, the speed of shocks are still 
determined by the right-hand side of \qref{E.ushock}.  

In order to deal with non-zero mean drift in initial data, 
we will use the following interesting
invariance of Burgers equation. 
Assume that $u_0(x)=o(|x|)$ as $|x| \to \infty$, and 
let $u(x,t)$ be the \CH\/ solution with $u(x,0)=u_0(x)$, 
defined for all $t \geq 0$.  Let $c\in\R$ and 
define
\be 
\label{E.Tc}
u^{(c)}_0(x) = u_0(x) + cx , \qquad 
T_c = \left\{ \begin{array}{rl} -c^{-1}, &\quad c <0, \\
+\infty, &\quad c \geq 0. \end{array} \right. 
\ee
Then the \CH\/ solution with initial data $u^{(c)}_0$ is given by
\be
\label{E.changedrift}
u^{(c)}(x,t) = \frac{1}{1+ct} u\left( \frac{x}{1+ct}, \frac{t}{1+ct} \right)
+ \frac{cx}{1+ct}, \quad t \in [0,T_c).
\ee
This is seen as follows. An elementary calculation shows that the
\CH\/ functionals for the different data are related by
\[ H^{(c)}(y,t;x) = H\left(y, \frac{t}{1+ct};\frac{x}{1+ct}\right) +
\frac{cx^2}{2(1+ct)}, \]
which implies the inverse Lagrangian functions are related by
\be
\label{E.changedrift2}
a^{(c)}(x,t) = a\left(\frac{x}{1+ct}, \frac{t}{1+ct}\right). 
\ee
We now substitute in \qref{E.hopf} to obtain \qref{E.changedrift}. 

\subsection{Solutions with \Levy\/ process initial data}\label{S.sol}
Here we describe how the solution of \qref{E.burgers},
with initial data of the form \qref{E.initial},
is determined in terms of Laplace exponents,
essentially following Bertoin's treatment in \cite{B_burgers}.

Suppose $x\mapsto u_0(x)$ is an arbitrary spectrally 
negative \Levy\/ process for $x\ge0$, with Laplace exponent $\psi_0$
having downward jump measure $\Lambda_0$. 
We first show that we may assume without loss of generality that
$\int_0^\infty (s\wedge s^2)\Lambda_0(ds)<\infty$.  Indeed, 
if $\int_1^\infty s\Lambda_0(ds)=\infty$, then
$u_0(x)/x \to-\infty$ almost surely as $x\to\infty$. (This follows from
the fact that for the compound Poisson process $X_x$ with jump measure
$\Lambda_0(ds) \one_{|s| \geq 1}$, one has $X_x/x\to\infty$ as $x\to\infty$
by the law of large numbers.)
In this case the \CH\/ function $H(y,t;x)$  has no minimum for any $t>0$, 
and equation \qref{E.burgers} has no finite entropy solution
for any positive time. 
Hence, we may suppose that $\int_1^\infty s\Lambda_0(ds)<\infty$. 

Next, we show that one may assume the mean drift $b_0=\E(u_0(1))$ is
zero. If $b_0$ is nonzero, we have $\lim_{x \to \infty} u_0(x)/x =
b_0$ a.s.\, by the strong law of large numbers.  
If $b_0<0$, then by comparison to compression-wave solutions 
with initial data $A+b \max(x,0)$, we find using the maximum
principle that a.s.\, the solution blows up exactly 
at time $-b_0^{-1}$. If $b_0>0$ there is a global solution. 
In either case, we may use the transformation
\qref{E.changedrift} with $c=b_0$ to reduce to the case $b_0=0$, replacing
$u_0(x)$ by $u_0(x)-b_0x\one_{x>0}$. More precisely, we apply
\qref{E.changedrift} for $x \geq 0$ noting that $a(0,t) \geq 0$,
thus $a(x,t) \geq a(0,t) \geq 0$ for $x \geq 0$, so that
\qref{E.changedrift2} holds for $x \geq 0$. 
We have:
\begin{lemma}
If $u^{(c)}_0$ is a spectrally negative \Levy\/ process with
\Levy\/ triplet $(c, \sigma^2_0, \Lambda_0)$, the Cole-Hopf
solution $u^{(c)}(x,t)$ is determined
via \qref{E.changedrift} for $x\ge0$ and $t \in  [0,T_c)$, 
in terms of a solution $u(x,t)$ having zero mean drift and 
defined for all $t\ge0$.
\end{lemma}

With these reductions, we may restrict ourselves to Laplace
exponents $\psi_0$ of the form 
\begin{equation}\label{psi0}
\psi_0(q) = \frac{\sigma_0^2q^2}2 + \int_0^\infty
\left(e^{-qs} - 1 + qs \right)
\Lambda_0(ds), \quad q\ge0.
\end{equation}
We will always assume that $a$ 
and $u$ are right continuous (i.e., $a(x,t)= a(x_+,t)$, compare with
\qref{E.ushock}). This ensures $a$ is an element of the 
Skorokhod space $\Skor$, so that we may use the 
standard Skorokhod topology to study limiting behavior.
For brevity, we write 
\[
v(x,t)=u(x,t)-u(0,t),\qquad 
l(x,t)=a(x,t)-a(0,t),
\]
and rewrite \qref{E.hopf} as 
\be
\la{E.vl}
 v(x,t) = \frac{x-l(x,t)}{t}, \quad x \geq 0,\ t > 0. 
\ee
Bertoin has shown that for all $t>0$, $x\mapsto l(x,t)$ is an increasing
\Levy\ process (a subordinator) with the same law as the first passage
process for $tu_0(x)+x$. 
We denote the Laplace exponents of  $l$ and $v$ by $\Phi$ and
$\psi$ respectively:
\be
\la{E.LElv}
\E\left(e^{-ql(x,t)}\right) = e^{-x\Phi(q,t)}, \quad \E
\left(e^{qv(x,t)}\right) = e^{x \psi(q,t)}, \quad x,q,t \geq 0.
\ee
We  combine (\ref{E.vl}) and \qref{E.LElv} to obtain 
\begin{equation}
\label{E.psi_t}
\psi(q,t) =\frac{q}{t}-\Phi\left(\frac{q}{t},t\right).
\end{equation}
Since $l$ is a subordinator, it has the simpler \LK\/ representation 
\begin{equation}
\label{E.phi}
\Phi(q,t) = d_tq + \int_0^\infty (1-e^{-qs})\mu_t(ds), \quad q >0,
\end{equation}
where $d_t \geq 0$ supplies the deterministic part of the drift, 
and $\mu_t$ is the \Levy\ measure  of $l(\cdot,t)$, 
which now must satisfy $\int_0^\infty (1\wedge s)\mu_t(ds)<\infty$ \cite{B_book}. 
We see from \qref{E.vl} that
$v(\cdot,t)$ is a \Levy\ process with no Gaussian component, 
and thus has a \LK\/ representation 
\be
\la{E.defpsi}
\psi(q,t) = b_t q + \int_0^\infty\left(e^{-qs} - 1 + qs \right)
\Lambda_t(ds), \quad q\ge0,\ t>0, 
\ee
related to \qref{E.phi} by
\be
\la{E.mulambda}
\Lambda_t( ds) =  \mu_t(t\,ds) ,
\qquad 
b_t + \int_0^\infty s\Lambda_t(ds) = \frac{1-d_t}{t}.
\ee

Due to the result that $l$ and the first passage process 
of $tu_0(x)+x$ have the same law, a simple functional relation holds between 
$\psi_0$ and $\Phi(q,t)$ \cite[Thm.~2]{B_burgers}:
\begin{equation}
\label{E.main}
\psi_0(t\Phi(q,t)) + \Phi(q,t)= q, \quad q \geq 0,\ t >0.
\end{equation}
The evolution takes a remarkably simple form when 
we combine equations (\ref{E.psi_t}) and (\ref{E.main}) to obtain
\begin{equation}
\label{E.psi_soln}
\psi(q,t) = \psi_0(q -t\psi(q,t)), \quad q \geq 0,\ t>0.
\end{equation}
But then $\psi(q,t)$ solves the inviscid Burgers equation (in $q$ and $t$!)
\begin{equation}
\label{E.psi_burgers}
\partial_t \psi + \psi \partial_q \psi =0, \quad \psi(0,q) = \psi_0(q).
\end{equation}
The solution to (\ref{E.psi_burgers}) may be constructed
by the method of characteristics and takes the form
(\ref{E.psi_soln}). The remarkable fact that the Laplace exponent is
also a solution to Burgers equation was first observed by Carraro and
Duchon~\cite[Thm.~2]{Duchon1}. 

Since $\psi_0$ is analytic and strictly convex,
the solution (\ref{E.psi_soln}) is analytic for all time and unique,
and the condition $\partial_q \psi(0,t)=0$ is preserved for all $t>0$. 
By \qref{E.psi_t}--\qref{E.mulambda}, we have
\begin{equation}
\label{E.phi_mass}
b_t = 0,  \qquad d_t = 1 - t \int_0^\infty 
s\Lambda_t(ds), \quad t>0.
\end{equation}
Let
\begin{equation}\la{E.m0}
M_0= \lim_{q \to \infty} \psi'_0(q) , \qquad
\end{equation}
We find $\Phi(q,t)\to\infty$ as $q\to\infty$ from (\ref{E.main}),
and differentiate to obtain 
\begin{equation}
\label{E.drift}
d_t = \lim_{q \to \infty} \partial_q \Phi(q,t) = \lim_{q\to \infty}
\frac{1}{1+t\psi_0'\left(t \Phi(q,t)\right)} = \frac{1}{1+tM_0}, \quad t >0,
\end{equation}
with the understanding that $d_t=0$ if $M_0=+\infty$. 
Then 
\begin{equation}\la{E.m}
M(t):= \lim_{q\to\infty} \D_q\psi(q,t) = \int_0^\infty s\Lambda_t(ds)
= \frac{M_0}{1+tM_0}, \quad t>0,
\end{equation}
with the understanding that $M(t)=t^{-1}$ when $M_0=\infty$. 
Note $M'=-M^2$. Below we will characterize the evolution
of $\Lambda_t$ differently.

\subsection{BV regularity}
It is clear from the \CH\ formula that 
$u$ is locally of bounded variation for every $t>0$. We derive a
decay estimate that quantifies this. 
The sample paths of $u_0$ have unbounded variation if and only
if~\cite[p.15]{B_book} 
\be
\label{E.sigma1}
\sigma_0^2 >0 \quad \mbox{or} \quad \int_0^\infty s \Lambda_0(ds) =
\infty.
\ee
Heuristically, this corresponds to the presence of many small
jumps (`dust'). This is reflected in the Laplace exponent as
$M_0=\lim_{q \to \infty} \psi_0'(q) = +\infty$ in this case. 
On the other hand, $M_0$ is finite if and only if 
$u_0$ is BV, in which case $\sigma_0=0$ and 
$M_0 = \int_0^\infty s \Lambda_0(ds) < \infty$.  

The analytic
formula (\ref{E.phi}) has the following probabilistic meaning. 
If we take a Poisson point process $x\mapsto m^t_x$ (masses of clusters) with
jump measure $\mu_t$ we have the representation~\cite[p.16]{B_book}
\begin{equation}
\label{E.a_repn}
l(x,t) = d_tx + \sum_{0 \leq y \leq x} m^{t}_y.
\end{equation}
The velocity field, and a point process of shock strengths 
$s^t_y= t^{-1}m^t_y$ are 
determined from  (\ref{E.hopf}), (\ref{E.drift}) and (\ref{E.a_repn}) by
\begin{equation}
\label{E.BV}
v(x,t) = M(t) x - \frac{1}{t}\sum_{0 \leq y \leq
  x} m^{t}_y  = M(t)x - \sum_{0 \leq y} s^t_y,
\end{equation}

For every $t>0$, $v(x,t)$ is the difference of
two increasing  functions: a linear drift and a pure jump process.
Thus, it is of bounded variation, and  by \qref{E.phi_mass} and
\qref{E.a_repn} we have  
\begin{equation}
\label{E.m1_finite}
\E \left( \int_0^x |\partial_y v(y,t)| dy \right) = 
2M(t)x ,
\quad x,t>0,
\end{equation}
because $\E \left( \sum_{0 \leq y \leq x} m^t_y\right) =
xM(t)=x(1-d_t)$. 

%
\subsection{Relation to Smoluchowski's coagulation equation}

We consider a positive measure $\nu_\tau(ds)$ interpreted as the 
number of clusters of mass or size $s$ per unit
volume at time $\tau$. Clusters of mass $r$ and $s$ coalesce by
binary collisions  at a rate governed by a symmetric kernel $K(r,s)$. 
A weak formulation of Smoluchowski's coagulation equation
can be based on a general moment identity 
for suitable test functions $\zeta$ (see~\cite{MP1}):
\begin{align}
\label{E.smol}
&\D_\tau \int_0^\infty \zeta(s)\,\nu_\tau(ds) 
 = \nonumber\\ &\qquad\qquad 
\frac{1}{2} \int_0^\infty \!\! \int_0^\infty
\left( \zeta(r+s)-\zeta(r)-\zeta(s) \right) K(r,s)\,\nu_\tau(dr) \,\nu_\tau(ds).
\end{align}
We consider only the {\em additive kernel \/}$K(r,s)=r+s$. It is classical that
(\ref{E.smol}) can then be solved by the Laplace
transform~\cite{Drake}. We denote the 
initial time by $\tau_0$ (to be chosen
below). The minimal (and natural) hypothesis on initial data
$\nu_{\tau_0}$ is that the mass $\int_0^\infty s \nu_{\tau_0}(ds)$ is
finite. We scale the initial data such that $\int_0^\infty s \nu_{\tau_0}
=1$. The Laplace exponent 
\begin{equation}\label{E.nu}
\vp(q,\tau) = \int_0^\infty
(1-e^{-qs}) \nu_\tau(ds)
\end{equation} 
then satisfies
\begin{equation}
\label{E.vp1}
\quad \partial_\tau \vp - \vp \partial_q \vp = -\vp, \quad \tau >\tau_0.
\end{equation}
We showed in~\cite{MP1} that (\ref{E.vp1}) may be used to define
unique, global, mass-preserving solutions to (\ref{E.smol}).
In particular, a map $\tau\mapsto \nu_\tau$ from $[\tau_0,\infty)$
to the space of positive Radon measures on $(0,\infty)$, such that
$\int_0^\infty s\nu_\tau(ds)=1$ for all $\tau\ge\tau_0$, is a solution
of Smoluchowski's equation with $K(r,s)=r+s$ (in an appropriate weak
sense detailed in  \cite{MP1}) if and only if $\vp$ satisfies \qref{E.vp1}.

We now connect solutions of the inviscid Burgers equation \qref{E.burgers}
with \Levy\/ process initial data to solutions of Smoluchowski's
equation through a change of scale.
Let $u_0$ satisfy \qref{E.initial} and assume as in subsection~\ref{S.sol}
that the corresponding downward jump measure $\Lambda_0$ satisfies 
$\int_0^\infty (s\wedge s^2)\Lambda_0(ds)<\infty$ and the mean drift is zero.
Let $\Lambda_t$ be the jump measure of the \CH\/ solution.
With $M_0$ and $M(t)$ as in \qref{E.m0} and \qref{E.m}, 
let $\tau_0=-\log M_0$ if $u_0$ is
of bounded variation, and $\tau_0= -\infty$ otherwise, and set
\be
\la{E.smolburgers}
\tau = -\log M(t), \quad 
\nu_\tau(ds) = \Lambda_t(M(t)ds).
\ee
From \qref{E.m} it follows $\int_0^\infty s \nu_\tau(ds)=1$,
and by \qref{E.nu} and \qref{E.defpsi} we find 
\be
\label{E.changevars}
\vp(q,\tau) = q - \psi(qe^\tau,t) .
\ee
We see that $\psi$ solves (\ref{E.psi_burgers}) if and only if $\vp$
solves \qref{E.vp1}. 
Therefore, the rescaled \Levy\/ measure of $v(\cdot,t)$
evolves according to Smoluchowski's equation.
Conversely, given any solution of Smoluchowski's equation with initial
data $\nu_0$ at a finite $\tau_0$, we can
construct a corresponding solution of \qref{E.burgers} by choosing
$u_0$ to be a spectrally negative \Levy\/ process with jump measure
$\Lambda_{t_0}$ via \qref{E.smolburgers}

Initial data $u_0$ with unbounded variation are  of particular 
interest. Here we have  {\em eternal solutions\/} $\nu_\tau$ to
(\ref{E.smol}) defined for all $\tau \in \R$. 
We see that eternal solutions are in
one-to-one correspondence with initial data $u_0$ of unbounded 
variation via \qref{E.changevars}. A finer correspondence
mapping the clustering of shocks to the additive coalescent is found
in~\cite{B_cluster}.  

To summarize, we have the following correspondence.

\begin{theorem}
\label{T.meanfield}
Assume $u_0$ is a spectrally negative \Levy\/ process with 
\Levy\/ triplet $(0, \sigma_0^2, \Lambda_0)$, with the same assumptions
as in Theorem~\ref{T.main}. 
Then for all $t>0$, $v(\cdot,t)$ and $l(\cdot,t)$ are 
  \Levy\/ processes with triplet $(0,0,\Lambda_t)$, whose jump
  measures $\Lambda_t$ determine a solution $\nu_\tau(ds)$ to Smoluchowski's coagulation
  equation with rate kernel $K(r,s)=r+s$ as described in
  \qref{E.smolburgers}. 
\end{theorem}


%
%
%
%
%
%
%
%
\subsection{Self-similar solutions}
Bertoin's characterization of eternal solutions is the analogue of the
\LK\ characterization of  infinitely divisible
distributions~\cite{B_book,Feller}. Among the latter, the 
stable distributions  are of particular interest, and their analogues
for Smoluchowski's equations are obtained by choosing the Laplace
exponent $\psi_0(q) = q^\alpha$, $\alpha \in (1,2]$. For $\alpha \in
(1,2)$ the corresponding \Levy\/ measures are 
\[
\Lambda(ds) =
\frac{ s^{-(1+\alpha)}}
{\Gamma(-\alpha)}
\,ds. 
\]
The Laplace exponent $q^2$
corresponds to an atom at the origin. We thereby
 obtain for $\alpha \in (1,2]$ a  family of self-similar 
solutions to Smoluchowski's equation with Laplace exponent of the form 
$\vp(\tau,q) = e^{-\beta \tau} \vp_\alpha(q e^{\beta \tau})$ where
$\vp_\alpha$ solves 
\begin{equation}
\label{E.add_scaling}
\vp_\alpha(q)^\alpha + \vp_\alpha(q) =
q, \quad q>0.
\end{equation}
The self-similar solutions to Smoluchowski's coagulation
equation are 
\begin{equation}
\label{E.nform}
\nu_\tau(ds) = e^{-2\tau/\beta} f_{\alpha}(e^{-\tau/\beta}s)\,ds, \quad
\beta=\frac{\alpha-1}{\alpha}, \quad \alpha \in (1,2],
\end{equation}
where $f_\alpha$ has been defined in \qref{E.ndensity}. 
An analytic proof that these are the only self-similar solutions to
Smoluchowski's equation may be found in~\cite{MP1}. Each of these
solutions corresponds to a self-similar process. Precisely,
let $X^\alpha$ denote the stable process with Laplace exponent
$q^\alpha$, and $T^\alpha$ and $V^\alpha$ denote the processes  
\begin{equation}
\label{E.first_pass}
T^\alpha_x=\inf \{y\geq 0: X^\alpha_y +y > x \}, \qquad
V^\alpha_{x} = x - T^\alpha_x.
\end{equation}
%
We have $M_0=+\infty$ and $M(t)=t^{-1}=e^{-\tau}$ in this case, and 
the Laplace exponent of the process $l(\cdot,t)$ is of the self-similar form
\begin{equation}
\label{E.ss3}
\Phi(q,t) = \vp(q,\tau)= t^{-1/\beta} \vp_\alpha \left( qt^{1/\beta}\right), \;\;
t >0. 
\end{equation}
The solution processes have the scaling property 
\begin{equation}
\label{E.shock_scaling}
l(x,t) \eqd t^{1/\beta} T^\alpha_{xt^{-1/\beta}}, 
  \qquad v(x,t) \eqd t^{1/\beta-1} V^\alpha_{xt^{-1/\beta}}.
\end{equation}
The corresponding \Levy\/ measures are obtained from
\qref{E.mulambda}, \qref{E.smolburgers} and \qref{E.nform}:
\be
\mu^\alpha_t(ds)= t^{-2/\beta} f_{\alpha}(t^{-1/\beta}s) \, ds,  \quad
\Lambda^{\alpha}_t(ds) = t^{1-2/\beta} f_{\alpha} \left(
t^{1-1/\beta} s \right) \, ds. 
\ee
In the important case $\alpha=2$, we have 
$1/\beta=2$ and $\vp_2(q)=-\frac12+\sqrt{\frac14+q}$, 
and by Laplace inversion~\cite[Ch.29]{Abramowitz}
we obtain the explicit expression in \qref{E.golovin2}
for the distribution of $T^2_x$.

\section{The convergence theorem}
\label{S.proof}
In \cite{MP1} we proved the following theorem characterizing solutions
that approach  self-similar form in Smoluchowski's coagulation
equation with additive  kernel. To every solution $\nu_\tau$ of
\qref{E.smol} with 
$\int_0^\infty s \nu_\tau(ds)=1$ we associate the probability
distribution function
\be\label{E.Fsd}
 F(s,\tau) = \int_{(0,s]} r\nu_\tau(dr). 
\ee
To a self-similar solution $f_{\alpha}$, $\alpha \in (1,2]$ with
  $\beta = (\alpha-1)/\alpha$ we associate
\begin{equation}
\label{E.Fdef}
F_{\alpha}(s) = \int_0^s rf_{\alpha}(r)\,dr = 
 \sum_{k=1}^\infty \frac{(-1)^{k-1}s^{k\beta}}{k!}
\Gamma(1+k-k\beta)\frac{\sin \pi k\beta}{\pi k\beta}.
\end{equation}
A probability distribution function $F^*$ is called nontrivial if
$F^*(s) <1 $ for some $s>0$; this means the distribution is proper
($\lim_{s \to \infty}F^*(s)=1$) and not concentrated at $0$.
\begin{theorem}
\label{T.add}
Suppose $\tau_1 \in \R$ and $\nu_{\tau}$, $\tau \in [\tau_1,\infty)$, is
  a solution to Smoluchowski's 
coagulation equation with additive kernel such that $\int_0^\infty s
\nu_{\tau_1}(ds) =1$.  
\begin{enumerate}
\item Suppose there is a rescaling function 
$\tlambda(\tau) \rightarrow \infty$ as $\tau\to\infty$ and a
nontrivial probability distribution function $F^*$ such that 
\begin{equation}
\label{E.add5}
\lim_{\tau \rightarrow \infty} F(\tlambda(\tau)s,\tau) = F^*(s) 
\end{equation}
at all points of continuity of $F^*$. Then there exists $\alpha \in (1,2]$ 
and a function $L$ slowly varying at infinity such that 
\begin{equation}
\label{E.add_M2}
\int_0^s r^2 \nu_{\tau_1}(dr) \sim {s^{2-\alpha} L(s)}
\quad\text{as $s \to \infty.$} 
\end{equation}
\item Conversely, assume that there exists $\alpha \in (1,2]$ and a
function $L$ slowly varying at infinity such that (\ref{E.add_M2})
  holds. Then there is a strictly increasing rescaling $\tlambda(\tau) \to
  \infty$ such that
\[ \lim_{\tau\to\infty} F(\tlambda(\tau) s,\tau) = F_{\alpha}(s), \quad 0
\leq s < \infty,\] 
where $F_{\alpha}$ is a distribution function for a self
similar solution as in \qref{E.Fdef}.
\end{enumerate}
\end{theorem}

We now prove Theorem~\ref{T.main}. Let $u_0$ be a spectrally
negative \Levy\/ process with zero mean drift and $\int_0^\infty (s \wedge
s^2) \Lambda_0(ds) < \infty$. To the solution increment
$v(x,t)=u(x,t)-u(0,t)$ with downward jump measure $\Lambda_t$, 
associate a solution $\nu_\tau$ of Smoluchowski's coagulation equation
\qref{E.smol} as in Theorem~\ref{T.meanfield}
with Laplace exponent $\vp(q,\tau)$ given by \qref{E.changevars}.
Let $\tau_1=\tau_0=-\log M_0$ if $M_0<\infty$, and
let $\tau_1=0$ if $M_0=+\infty$ and $\tau_0=-\infty$.

We deduce Theorem~\ref{T.main} from Theorem~\ref{T.add} by 
establishing two equivalences:
\begin{itemize}
\item[(a)] 
There is a rescaling  $\lambda(t) \rightarrow \infty$ as $t\to\infty$
and a \Levy\ process $V^*$ with zero mean drift $\E(V^*)$ such that
$\Vt \law V^*$ 
if and only if 
there is a rescaling $\tlambda(\tau)\to\infty$ as $\tau\to\infty$
and a nontrivial probability distribution function $F^*$
such that \qref{E.add5} holds.
\item[(b)] 
$ \int_0^\infty s^2 \nu_{\tau_1}(ds)< \infty$ if and only if 
$\int_0^\infty s^2   \Lambda_0 (ds) < \infty$. 
Moreover, 
$\int_0^s r^2 \nu_{\tau_1}(dr) \sim s^{2-\alpha}L(s)$ as $s \to \infty$ 
if and only if 
$\int_0^s r^2 \Lambda_0(dr) \sim s^{2-\alpha}L(s)$ as $s \to \infty$. 
\end{itemize}

{\em Proof of (a).}
We prove claim (a) by showing each part equivalent
to a corresponding convergence statement for rescaled Laplace
exponents.  
First, convergence in law in
$\Skor$ for  processes with independent 
increments can be reduced to the convergence of characteristic
exponents~\cite[Cor. VII.4.43, p.440]{Jacod}.
In particular, suppose $\lambda(t)\to\infty$ as $t\to\infty$.
Then we have 
\be
\la{E.eqv1}
\Vt \law V^*,  \quad\mbox{with \ $\E(V^*_1)=0$,} 
\ee
if and only if
$\E(e^{ik\Vt_x}) \to \E(e^{ikV^*_x})$ for all 
{$k \in \R$, uniformly for $x$ in compact sets, and $\E(V^*_1)=0$}.
But since we are working with \Levy\/ processes, the \LK\/ formula
shows the dependence on $x$ is trivial, and thus \qref{E.eqv1} is equivalent to 
\be
\la{E.eqv2}
\E(e^{ik\Vt_1}) \to \E(e^{ikV^*_1}) \quad\mbox{for all $k \in \R$, 
\quad and\ \ $\E(V^*_1)=0$}.
\ee
But pointwise convergence of characteristic functions is equivalent to
convergence in distribution of the random variables
$\Vt_1$~\cite[XV.3.2]{Feller}, and since $\Vt_1=1-\Tt_1 
\leq 1$, \qref{E.eqv2} is equivalent to convergence of the Laplace
transforms~\cite[XIII.1.2]{Feller}:
\be
\la{E.eqv4}
\E(e^{q\Vt_1}) \to \E(e^{qV^*_1}) \quad \mbox{for all $q >0$}, 
\quad\mbox{and}\ \  \E(V^*_1)=0. 
\ee
Taking logarithms and using \qref{E.rscV} and \qref{E.LElv}, 
\qref{E.eqv4} is equivalent to 
\be \la{C.psi}
\lambda \psi\left({qt}/{\lambda},t\right) \to \psi^*(q) 
\quad\mbox{for all $q>0$}, 
\quad \mbox{and}\ \ \D_q\psi^*(0)=0,
\ee
where $\E(e^{qV^*_x})=e^{x\psi^*(q)}$.
This expresses the convergence of $\Vt$ in terms of convergence
of rescaled Laplace exponents.

Now suppose $\tlambda(\tau)\to\infty$ as $\tau\to\infty$. 
Using~\cite[XIII.1.2]{Feller} again, the (proper) convergence in \qref{E.add5}
is equivalent to pointwise convergence of Laplace transforms:
\be \la{C.eta}
\eta(q,\tau) \to \eta^*(q)  \quad\mbox{for all $q>0$},
\quad\mbox{with}\ \ \eta^*(0)=1.
\ee
where $\eta(q,\tau) := \int_0^\infty e^{-qs}
F(\tlambda(\tau)\,ds,\tau)$,
$ \eta^*(q) := \int_0^\infty e^{-qs} F^*(ds)$.
By \qref{E.Fsd} and \qref{E.nu}, we have
\be 
\eta(q,\tau)= (\D_q\vp)(q/\tlambda,\tau), \qquad
\int_0^q \eta(r,\tau)\,dr = \tlambda \vp(q/\tlambda,\tau).
\ee
We claim that \qref{C.eta} is equivalent to the statement that
(with $\vp^*(q)=\int_0^q \eta^*(r)\,dr$)
\be \label{C.vp}
\tlambda\vp(q/\tlambda,\tau) \to \vp^*(q) 
\mbox{\ \ for all $q>0$}, \quad\mbox{and}\quad
\D_q\vp^*(0)=1.
\ee
Clearly, since $\eta(\cdot,\tau)$ is completely monotone
and bounded, \qref{C.eta} implies \qref{C.vp}. In the other
direction, assume \qref{C.vp}. For any sequence $\tau_j\to\infty$
there is a subsequence along which $\eta(q,\tau_j)$ converges
for all (rational, hence real) $q>0$, to some limit whose integral
must be $\vp^*$. Thus \qref{C.eta} follows.

We now finish the proof of claim (a) by observing that due
to \qref{E.changevars}, we have
\be
\tlambda\vp(q/\tlambda,\tau)= q - 
\tlambda\psi({q e^\tau}/{\tlambda},t).
\ee
Hence the convergence in \qref{C.psi}
is equivalent to that in \qref{C.vp} provided we have
\be \la{E.lrel}
\lambda(t)/t = \tlambda(\tau)/e^\tau, 
\ee
or $\lambda(t) = tM(t) \tlambda(\tau)$, since $tM(t)\to1$ as $t\to\infty$. 
(Note $tM(t)=1$ if $M_0=\infty$.)

%
%
%
{\em Proof of (b).}
It is only the case $M_0=\infty$ that requires some work. Indeed, if
$M_0 < \infty$ we see from \qref{E.smolburgers} that $\nu_{\tau_1}(ds)
= \Lambda_0(M_0 \,ds)$. In what follows, we suppose that $M_0=\infty$.
We then have an eternal solution to Smoluchowski's equation, and $t=e^\tau$. 
We shall compare the tails of
$\nu_{0}$ ($\tau=0$) with that of $\Lambda_0$ ($t=0$). 

Claim (b) is a purely analytic fact that follows from Karamata's Tauberian
 theorem~\cite{Feller}. We first reformulate it as a  statement about
 Laplace transforms.  Let $\vp_0(q)=\vp(q,0)$, $\psi_0(q)=\psi(q,0)$.
For every $\alpha \in (1,2]$ we have
\begin{equation}
\label{eq:tauber1}
\int_0^s r^2 \nu_0(dr) \sim s^{2-\alpha}L(s) \iff 1-\vp_0'(q) \sim
q^{\alpha-1} L\left(\frac{1}{q}\right) \frac{\Gamma(3-\alpha)}{\alpha-1}
\end{equation}
as $s \to \infty$ and $q \to 0$ respectively (see~\cite[eq. 7.4]{MP1}). 
By the same argument,
\begin{equation}
\label{eq:tauber2}
\int_0^s r^2 \Lambda_0(dr) \sim s^{2-\alpha}L(s) \iff \psi_0'(q) \sim
q^{\alpha-1} L\left(\frac{1}{q}\right) \frac{\Gamma(3-\alpha)}{\alpha-1},
\end{equation}
with the following caveat when $\alpha=2$. 
If $\int_0^\infty r^2 \Lambda(dr)=\infty$ then
(\ref{eq:tauber2}) holds. On the other hand, if $\int_0^\infty r^2
\Lambda_0(dr) <\infty$ then we must modify the second condition
in (\ref{eq:tauber2}) to
\[\psi_0'(q) \sim \left( \sigma^2 + \int_0^\infty r^2 \Lambda(dr)\right) q,
\quad q \rightarrow 0.\]
We set $t=1$ in (\ref{E.changevars}) and differentiate
\qref{E.psi_soln} with respect to $q$ to obtain 
\begin{equation}
\label{eq:psiphi}
\psi_0'(\vp_0(q)) = \frac{1- \vp_0'(q)}{\vp_0'(q)} 
= \frac1{\vp_0'(q)}-1.
\end{equation}
The functions $\psi_0'$, $\vp_0$, and $1/\vp'$ 
are strictly increasing. Since $\vp_0'(0)=1$
we also have $\vp_0(q)= q(1+o(1))$ as $q \to 0$. A sandwich
argument as in~\cite{Feller} may now be used to deduce claim (b). 
First suppose that (\ref{eq:tauber1}) holds. Fix $b,\ve>0$. 
Then for $q$ sufficiently small we use monotonicity
and (\ref{eq:psiphi}) to obtain
\[
\frac{1- \vp_0'(bq(1-\ve))}{1-\vp_0'(q(1+\ve))} 
\frac{\vp_0'(q(1+\ve))}{\vp_0'(bq(1-\ve))}
<
\frac{\psi_0'(bq)}{\psi_0'(q)}  < 
\frac{1- \vp_0'(bq(1+\ve))}{1-\vp_0'(q(1-\ve))}
\frac{\vp_0'(q(1-\ve))}{\vp_0'(bq(1+\ve))}.
\]
Letting first $q$ and then $\ve \to 0$, we obtain 
\[ \lim_{q \to 0} \frac{\psi_0'(bq)}{\psi_0'(q)} = b^{\alpha-1}.\]
Thus, $\psi_0'$ is regularly varying with exponent $\alpha-1$.
Similarly, if we assume that (\ref{eq:tauber2}) holds, we sandwich
\[ 
\frac{\psi_0'(b(1-\ve)q)}{\psi_0'((1+\ve)q)} <
\frac{1-\vp_0'(bq)}{1-\vp_0'(q)} 
\frac{\vp_0'(q)}{\vp_0'(bq)}.
<
\frac{\psi_0'(b(1+\ve)q)}{\psi_0'((1-\ve)q)},
\]
to deduce that $1-\vp_0'$ is regularly varying with exponent
$\alpha-1$. Finally, since $\vp_0'(0)=1$ it follows from (\ref{eq:psiphi}) that 
$\lim_{q \to 0} {\psi_0'(q)}/(1-\vp_0'(q))=1.$
This finishes the proof of Theorem~1.

\section{Energy, dissipation and spectra}
\label{S.energy}
In this section, we compute several statistics of physical
interest for the solution increments: mean energy and dissipation, 
the law of the Fourier-Laplace transform, and the multifractal spectrum.
While the computations are routine, some
interesting features emerge, namely (i) conservation of energy
despite dissipation at shocks, (ii) a simple evolution rule for the
Fourier-Laplace spectrum, and (iii) a multifractal spectrum in sharp
variance with 
that of fully developed turbulence.  Simple proofs of fine regularity
properties (e.g., Hausdorff dimension of the set of Lagrangian regular
points) may be found in~\cite{B_burgers}. 

\subsection{Energy and dissipation}
The energy in any finite interval $I \subset \R_+$is computed using
the \LK\/ formula \qref{E.defpsi} and Fubini's theorem as follows. 
\ba 
\nn
&& \E\left(\int_I v(x,t)^2 \, dx \right) = 
\int_I \E \left( v(x,t)^2\right) \, dx =
\int_I
\left( \partial_q^2\left. \E \left( e^{q v(x,t)}\right) 
 \right|_{q=0} \right) dx \\
\nn 
&&
= \int_I \left. \partial_q^2 e^{x \psi(q,t)} \right|_{q=0} \, dx 
= \int_I \left( x^2 \partial_q \psi (0,t)^2 + x \partial_q^2 \psi(0,t)
\right) \, dx \\
\label{E.energy}
&& 
= b_t^2 \int_I x^2\, dx  + \left(\int_0^\infty y^2 \Lambda_t(dy)
\right) \int_I x\,dx. 
\ea
Let us restrict attention to solutions of mean zero, that is
$b_t=0$. Then we have conservation of energy in the sense that 
\be
\label{E.cons2}
\E \left(\int_I v(x,t)^2 \, dx\right)= \E \left( \int_I v(x,0)^2 \, dx
\right), \quad t \geq 0. 
\ee
Indeed, by \qref{E.energy}, we see that \qref{E.cons2} is equivalent to
\be
\label{E.cons3}
 \partial_q^2 \psi(0,t) =  \partial_q^2 \psi_0 = \sigma_0^2 +
\int_0^\infty s^2 \Lambda_0(ds) =: M_2, 
\ee
with the understanding that $\partial_q^2\psi(0,t) =\infty$ if
$\int_0^\infty s^2 \Lambda_0(ds)$ is divergent. It is only necessary
to differentiate \qref{E.psi_soln} to obtain
\[ \partial_q \psi(q,t) = \frac{\psi_0'(q-t\psi)}{1+ t \psi_0'(q-t\psi)}, \quad 
\partial_q^2 \psi(q,t) = \frac{1}{\left(1+
  t\psi_0'(q-t\psi)\right)^3} \psi_0''(q-t\psi), \]
and then take the limit $q\to 0$ to obtain \qref{E.cons2}. 

The dissipation at a shock with left and right limits $u_\pm$ is
obtained as follows. The decay 
of the $L^2$ norm for solutions to Burgers equations with viscosity
$\eps$,  $u_t + uu_x = \eps u_{xx}$, is given by
\[ \frac{d}{dt} \int_{\R} u^2 \, dx = 2 \eps \int_{\R} u_x^2 \, dx. \]
The right hand side may be evaluated exactly for traveling waves (viscous
shocks) of the form $u(x,t)=u^\eps(x-ct)$. It is easily seen that for
any $\eps>0$ a traveling wave profile connecting the states $u_- >
u_+$ at $\mp \infty$ is of the form $u^\eps(x-ct)= w((x-ct)/\eps)$
where $w$ satisfies the  ordinary differential equation
\[ -c\left(w-u_-\right) + \frac{1}{2}\left(w^2 -u_-^2 \right)  =
\frac{dw}{d\xi},
\quad c = \frac{u_-+u_+}{2}.\]  
We therefore have 
\ba
\nn
&& 2 \eps \int_\R u_x^2 \, dx = 2 \int_\R (w')^2 \, d\xi =2
  \int_{u_-}^{u_+} \left[-c(w-u_-) +  \frac{1}{2}\left(w^2 -u_-^2 
  \right) \, \right] dw\\ 
\nn
&& 
= 2(u_--u_+)^3 \int_0^1 w(1-w) \, dw  = \frac{(u_--u_+)^3}{3}. 
\ea
The right hand side is independent of $\eps$ and captures the
dissipation of the entropy solution in  the limit $\eps \to 0$. 
The dissipation at shocks in any finite interval $I \subset \R_+$ may now be
computed by summing over all shocks in $I$ using \qref{E.a_repn} and
\qref{E.BV}:
\be
\label{E.diss}
\frac{1}{3}\E \left( \sum_{y \in I} (v(y_-,t)- v(y_+,t))^3\right)
= \frac{1}{3} \E \left( \sum_{y \in I} (s_y^t)^3 \right) 
= \frac{|I|}{3} \int_0^\infty s^3 \Lambda_t(ds), 
\ee
where $|I|$ is the length of $I$.

Conservation of energy in the sense described in \qref{E.cons2} is
rather surprising in 
view of the dissipation  at shocks. In particular, there are
solutions with finite energy ($\int_0^\infty s^2 \Lambda_t(ds) < \infty$),
but infinite dissipation ($\int_0^\infty s^3 \Lambda_t(ds) =
\infty$). However, there is no contradiction, since \qref{E.cons2}
refers to the expected value of the energy in any finite interval $I$, and
the energy dissipated in shocks is compensated by energy input
from the endpoints of $I$. 

\subsection{The Fourier-Laplace spectrum}
We show that the law of the Fourier transform $\hat{v}(k,t)$ 
of paths $x\mapsto v(x,t)$, is determined by a \Levy\/ process
with jump measure $s^{-1}\Ltail_t(s)\, ds$, where 
$\Ltail_t(s)= \int_s^\infty \Lambda_t(ds)$. 
Here $\Lambda_t(ds)$ denotes the jump measure of $v(x,t)$
(see Theorem~\ref{T.spectrum} below). 
The assertion $\hat{v} \sim k^{-1}$ as $k \to \infty$ 
for white noise initial data is common in the Burgers turbulence
literature (e.g., see \cite{Frisch-Parisi,Woc}).    
For the present case of \Levy\/ process initial data, 
we show that $\hat{v}(k,t) \sim -iM(t)k^{-2}$ as $k \to \infty$. 
In addition, we find precise corrections under additional assumptions on
$\Lambda_t$ (for example, for self-similar solutions). 

These computations with the laws of the Fourier-Laplace
transform should be contrasted with the conventional notion of 
the power spectrum.
Despite its widespread use for wide-sense stationary processes,
the power spectrum is of limited utility for the present problem 
involving stationary increments, as we now show.
Fix $L>0$ and consider the interval $[0,L]$. Almost every sample path 
$v(x,t)$ is bounded on $[0,L]$ and we may define the truncated Fourier
transform
\be
\label{E.Fourier1}
 \hvl (k, t) = 
\int_0^L e^{-ikx} v(x,t) \, dx. 
\ee
If the energy is finite ($M_2 < \infty$ in \qref{E.cons3}), we may
compute a truncated power spectral density $S_L(k)$ as follows. We have
\be
\label{E.Fourier2}
\frac1L |\hvl(k,t)|^2 = \frac{1}{L} \int_0^L \int_0^L e^{-ik(x-y)} v(x,t)
v(y,t) \, dx \, dy. 
\ee
Since $v(x,t)$ is a \Levy\/ process with mean zero, 
the autocorrelation is
\be
\mathbb{E}(v(x,t) v(y,t)) = (x\wedge y)M_2. 
\ee
We take expectations in \qref{E.Fourier2} to find 
\be
\label{E.Fourier3}
S_L(k):= \frac1L \E\left(|\hvl(k,t)|^2\right) = \frac{2 M_2}{k^2}
\left(1-\frac{\sin kL}{kL}\right),\quad k \neq 0.
\ee
The power spectrum $S(k)=\lim_{L\to\infty}S_L(k)=2M_2/k^2$ is now seen to be 
well-defined, but is unsuitable for distinguishing solutions 
because all solutions with
the same energy (possibly infinite) have identical power spectrum. 

A well-defined spectrum that distinguishes solutions may be obtained by 
taking the Fourier-Laplace transform of process paths. 
For fixed $p>0$ we define the random variable
\be
\label{E.Fourier4}
\lap{v}(p,t) = \int_0^\infty e^{-px} v(x,t) \,dx =
\frac{1}{p}\int_0^{\infty} e^{-px} v(dx,t). 
\ee
The integrals are well-defined because $\lim_{x\to \infty}
v(x,t)/x=0$ a.s.\ by the strong law of large numbers, and $v(x,t)$ is of
bounded variation. If $s^t$ denotes a point process of shock strengths
as in \qref{E.BV} we have 
\be
\label{E.Fourier_CM}
p\lap{v}(p,t) = \frac{M(t)}{p} - \sum_{0\le x} e^{-px}s^t_x.
\ee
We determine the law of $p\lap{v}(p,t)$ by computing its Laplace transform 
via the `infinitesimal' \LK\/ formula  $\E\left( e^{q v(dx,t)} \right) =
e^{\psi(q,t) \, dx}$. We then have
\ba
\nn
\lefteqn{\E \left(e^{q p \lap{v}(p,t)}\right) = 
\exp \left( \int_0^\infty \psi( q e^{-px}, t) \, dx \right)} \\
\label{E.Fourier5}
&& = \exp \left( \frac{1}{p} \int_0^q \frac{\psi(q',t)}{q'} \,
dq'\right) =: \exp \left( {\frac{1}{p}\pss(q,t)}\right), \quad p,q>0,
\ea
after the change of variables $q'=q e^{-px}$. We now observe that 
$\pss$ determines a Laplace exponent as follows. 
Let $\Ltail_t(s) = \int_s^\infty \Lambda_t(ds)$ denote the tail of the 
\Levy\ measure $\Lambda_t$. Since $\int_0^\infty (s\wedge s^2)
\Lambda_t(ds) < \infty$ we have the bounds  
\[ s\Ltail_t(s) \leq \int_s^\infty r \Lambda_t(dr),
\qquad s^2 \Ltail_t(s) \leq \int_0^\eps r^2 \Lambda_t(dr)
+ s^2 \Ltail_t(\eps), \quad s \in (0,\eps). \]
Therefore, 
\[
\lim_{s \to \infty} s \Ltail_t(s) = 0, 
\qquad \lim_{s \to 0} s^2 \Ltail_t(s)=0, 
\]
and we may integrate by parts in \qref{E.defpsi} to obtain
\be
\la{E.pss1}
\frac{\psi(q',t)}{q'} 
= \int_0^\infty (1-e^{-q's}) \Ltail_t(s) \, ds.
\ee
Integrating once more in $q'$ we find
\be
\label{E.Fourier6}
\pss(q,t) = 
\int_0^q \frac{\psi(q',t)}{q'} \,dq' = 
\int_0^\infty (e^{-qs}-1 + qs) \frac{\Ltail_t(s)}{s} \,
ds. 
\ee
We integrate by parts in \qref{E.m} to see that 
\be
\label{E.Lam1}
\int_0^\infty \Ltail_t(s) \, ds= \int_0^\infty s
\Lambda_t(ds) =  M(t) <\infty.
\ee
This enables us to write
\be
\label{E.phss1}
\pss(q,t) = M(t)q - \phss(q,t), \quad 
\phss(q,t) = \int_0^\infty (1-e^{-qs}) \frac{\Ltail_t(s)}{s} \, ds.
\ee
Since \qref{E.Lam1} ensures $s^{-1}\Ltail_t(s) \,ds$ satisfies 
the finiteness conditions for a jump measure, $\pss$ is a
Laplace exponent for a \Levy\/ process with zero mean drift that we denote
by $Z^t$. Similarly, $\phss$ is the Laplace exponent for a subordinator that
we denote $Y^t$.  We summarize our calculations in the identities
\be
\label{E.phss2}
Z^t_r  =  M(t)r -Y^t_r, \;\;
\E\left(e^{qZ^t_r} \right) = e^{r \pss(q,t)}, \;\; \E\left(e^{-qY^t_r}
\right) = e^{-r \phss(q,t)},\;\; r,q, t >0.
\ee

The result is that the Laplace spectrum of the solution increments is determined by
\be
\label{E.Fourier7}
\E \left(e^{q p \lap{v}(p,t)}\right) = \E \left( e^{ q Z^t_{1/p}} \right),
\quad q>0,\ p >0,
\ee
which implies that $\lap{v}(p,t)$ has the same law as $p^{-1} Z_{1/p}^t$
for fixed  $p>0$. Note that $\lap{v}(p,t)$ is not a \Levy\/ process
in $p$. In fact, for a fixed realization, $\lap{v}(p,t)$ is analytic
in $p$. Nevertheless, its law is determined by the \Levy\/ process $Z^t$. 

We extend this computation to the Fourier spectrum ($p=ik$) as
follows. The calculations leading to \qref{E.Fourier5} hold for
complex $q$ with $\mathrm{Re}(q) \geq 0$, and in particular for $q =
i\xi$, $\xi \in \R$. Moreover, $\lap{v}(p,t)$ is a well-defined random
variable for every $p$ with $\mathrm{Re}(p)>0$. Thus, we may
analytically continue the identity 
$E(e^{qp \lap{v}(p,t)}) = \exp( p^{-1}\pss(q,t))$  
to  all $p$ with $\mathrm{Re}(p)>0$, and
$q=i\xi$. As in \qref{E.LKu}, let $\Pss(\xi,t)=-\pss(i\xi,t)$ define
the characteristic exponent corresponding to the \Levy\/ process $Z^t$.
For $\eps, k>0$ we set $p= \eps +ik$, $\hat{v}(k-i\eps,t)=\lap{v}(\eps+ik,t)$ 
and pass to the limit $\eps \to 0$
on both sides of \qref{E.Fourier5} to obtain
\ba
\lim_{\eps \dnto 0} \lefteqn{\E \left(e^{i\xi (ik \hat{v}(k-i\eps,t))}
\right) = \exp \left( \frac{1}{ik}\pss(i\xi,t) \right) } \\
&& = \exp \left( \frac{i}{k}\Pss(\xi,t) \right) = \E \left( e^{ i\xi
    Z^t_{1/k}} \right), \quad  \xi \in \R, k>0. 
\ea
Thus, for fixed $k >0$, as $\eps\dnto 0$ 
the random variables $ik\hat{v}(k-i\eps,t)$ converge in law 
to the (real) random variable $Z^t_{1/k}$. 
We denote this limit by $ik\vfour(k,t)$. 
As before we do not assert that
the processes $ik\hat{v}(k-i\eps,t)$ converge in law to the process
$Z^t_{1/k}$, simply the convergence of random variables for fixed
$k$. We summarize our conclusions as follows.
\begin{theorem}
\label{T.spectrum}
Let $Y^t$ be a subordinator with Laplace exponent $\phss(q,t)$ from 
\qref{E.phss1}, and let $Z^t$ the \Levy\/ process defined by \qref{E.phss2}. 
Then  for every fixed $p>0$ and $k>0$ the random variables $p\lap{v}(p,t)$ 
and $ik \vfour(k,t)$ have the same law as $Z^t_{1/p}$ and $Z^t_{1/k}$,
respectively.
\end{theorem} 

Due to this result and \qref{E.phss2},
we always have the upper bound $ik^2\vfour(k,t) \leq  M(t)$ a.s. 
This crude bound may be refined as $k \to \infty$ using information
related to the sample path
behavior of subordinators (see~\cite[Ch.~III.4]{B_book}). 
\begin{corollary}
\label{C.asympt1}
For every $t>0$, $\lim_{k \to \infty} ik^2 \vfour(k,t) = M(t)$ in probability. 
\end{corollary}
\begin{proof}
This follows from the fact that 
$\lim_{r \dnto 0}Y^t_r/r=0$ in probability, proved as
follows. By \qref{E.phss2} we have
$\E(e^{-q Y^t_r/r})= e^{-r\Phi_\#(q/r,t)}$, and since
$1-e^{-s}\le 1\wedge s$, by \qref{E.phss1} we have
that the Laplace exponent 
\[
r\Phi_\#(q/r,t) = r\int_0^\infty(1-e^{-qs/r})
\frac{\Ltail_t(s)}s\,ds
\le  \int_0^\infty (r\wedge qs)
\frac{\Ltail_t(s)}s\,ds \to 0
\]
as $r\dnto0$ for each $q>0$.
Hence $Y^t_r/r\to 0$ in law.
\end{proof}

A similar conclusion holds for the Laplace spectrum as $p\to\infty$.
Actually, for the subordinator $Y^t_r$, the sample paths have the stronger
property that $\lim_{r\dnto0} Y^t_r/r\to 0$ a.s.~\cite[III.4.8]{B_book}.
Under a mild assumption on the integrability of the small jumps, 
we can strengthen convergence in probability to almost-sure
convergence of the Laplace spectrum. 

\begin{corollary}
\label{C.asympt2}
For every $t>0$, $\lim_{p\to\infty} p^2\lap{v}(p,t)=M(t)$ in
probability.  If we also assume $\int_0^1 |\log s| \Ltail_t(s)\,ds < \infty$, 
then $\lim_{p \to \infty} p^2 \lap{v}(p,t)=M(t)$ a.s.
\end{corollary}
\begin{proof}
For notational convenience, we suppress the dependence on $t$ in the proof.
Fix $\eps >0$, 
and let $p_m= 2^m$ for positive integers $m$. We will show that
$\lim_{m \to \infty} p_m^2\lap{v}(p_m) =M$ a.s. That is, for every
$\eps>0$, we claim 
\be
\label{E.bc1}
\prob\left( \left| p_m^2 \lap{v}(p_m) - M \right| > \eps 
\ \mbox{infinitely often}\right) =0.
\ee
This is sufficient to establish $\lim_{p\to \infty}p^2 \lap{v}(p)=M$
a.s. Indeed, since $M/p- p\lap{v}(p)$ is completely monotone by
\qref{E.Fourier_CM}, for  $p\in (p_m,p_{m+1})$ we have the bounds
\[ 
0 < M(t) - p^2\lap{v}(p) 
< 
\frac{p}{p_m} \left( M(t) - p^2_{m}\lap{v}(p_{m}) \right)
<
 2 \left( M- p_m^2 \lap{v}(p_m)\right) ,
\]
and therefore 
\be
\{M-p^2\lap{v}(p) > 2\eps\} \subset \{ M - p_m^2 \lap{v}(p_m) > \eps \}.
\ee
%

In order to prove \qref{E.bc1} we use the elementary estimate
\ba
\nn
&& \prob \left( \left| p_m^2 \lap{v}(p) - M \right| > \eps \right)
= \prob\left( p_m Y_{1/p_m} > \eps \right) 
\leq \frac{e}{e-1}\E\left(1 -  \exp \left(-\frac{p_m}{\eps}
      Y_{1/p_m}\right) \right)  \\
\nn
&& = \frac{e}{e-1} \left(1-
  \exp\left(-\frac{1}{p_m}\phss(\frac{p_m}{\eps}) \right) \right) 
\leq \frac{e}{e-1}
\frac{1}{p_m}\phss(\frac{p_m}{\eps}).
\ea
We will show that $\sum_{m=1}^\infty p_m^{-1} \phss(p_m/\eps) <
\infty$. The first Borel-Cantelli lemma then implies \qref{E.bc1}. 
For clarity, we suppose $\eps=1$. This causes no essential
difference and reveals the main computation.

Denote the integrated tail of the \Levy\ measure for $Y^t$ by
\be 
\label{E.pibar}
\pibar_t(s) = \int_s^\infty \frac{\Ltail_t(s')}{s'} \, ds'.
\ee 
We integrate by parts and use Tonelli's theorem to find
\be
\label{E.lil3}
\sum_{m=1}^\infty p_m^{-1} \phss(p_m) = \int_0^\infty
\sum_{m=1}^\infty e^{-p_m s} \pibar_t(s) \,ds. 
\ee
It is only necessary to check that the integral over $s\in (0,1)$ is
finite. Here we use the elementary estimate 
\[ 
\sum_{m=1}^\infty e^{-2^m s} \leq \int_0^\infty \exp(-e^{x\log 2}s)
\,dx  = \frac{1}{\log 2} \int_{s}^\infty e^{-y} \frac{dy}{y} \leq
\frac{|\log s|+1}{\log 2},
\]
so that
\[ 
\int_0^1 \sum_{m=1}^\infty e^{-p_m s} \pibar_t(s) \,ds \leq
\frac{1}{\log 2} \int_0^1 (1+|\log s|) \pibar_t(s) \,ds. 
\] 
By the definition of $\pibar_t(s)$ in \qref{E.pibar}, the last integral is
\ba
\nn
&& {\int_0^1 |\log s| \int_s^\infty \frac{\Ltail_t(r)}{r} \, dr \, ds 
= \int_0^\infty \frac{\Ltail_t(r)}{r} \, dr \int_0^{1\wedge r}
|\log s| \, ds} \\
\nn
&&\qquad \leq \int_0^1 |\log r| \Ltail_t(r)\, dr + \int_0^\infty
\Ltail_t(r)\, dr, 
\ea
which is finite by assumption. 
\end{proof}

Corrections to the bound $ik^2\hat{v}(k,t)\le M(t)$ 
involve the law of the iterated logarithm~\cite[III.4]{B_book}. 
The following corollary holds for initial data that is not 
BV (so $M_0=+\infty$ and $M(t)=1/t$) with suitably regular small
jumps (`dust'). 
\begin{corollary}
\label{C.LIL}
Assume $\sigma_0\ne0$ and $\alpha=2$, or assume 
$\sigma_0=0$ and $\Ltail_0(s)=\int_s^\infty \Lambda(dr)$ is regularly varying at 
zero with exponent $-\alpha$ where $\alpha \in (1,2)$. 
Then for every $c>0$ and $t>0$ we have 
\be
\label{E.LIL7} 
 -\log \prob\left( \frac{t^{-1}- ik^2\vfour(k,t)}
{h(k\log\log k)} \leq c \right) \sim
\frac  
{\log\log k}
{\gamma c^\gamma t^{1+2\gamma}}
, \quad k \to \infty,
\ee
where $\gamma=1/(\alpha-1)$ and $h(k)=k/\psi_0(k)$.
\end{corollary}

This corollary is a consequence of~\cite[Lemma III.12]{B_book} and is
associated with the following lemma of independent interest which shows that
the evolution preserves the regularity of the dust. 
\begin{lemma}
\label{L.LIL}
(a) Assume that $\sigma_0=0$ and $\Ltail_0(s)$ is regularly varying at
zero with exponent $-\alpha$, $\alpha \in (1,2)$. Then $\Ltail_t(s)$ is
regularly varying at zero with exponent $-1/\alpha$ for every $t>0$.

(b) If $\sigma_0\neq 0$, then $\Ltail_t(s) \sim (\sigma_0 t)^{-1}
\sqrt{2/(\pi s)}$ as $s \to 0$, for every $t>0$. 
\end{lemma}
\begin{proof}
Recall that $t^{-1}-ik^2\hat{v}(k,t)$ agrees in law with
$kY^t_{1/k}$.
Combining \qref{E.psi_t} with \qref{E.Fourier6} we find
that the Laplace exponent of the subordinator $Y^t$ satisfies
\be\label{E.Phish}
\Phi_\#(q,t) = \int_0^q \Phi\left(\frac{q'}{t},t\right) \frac{dq'}{q'}
= \int_0^{q/t} \Phi\left({q'},t\right) \frac{dq'}{q'}.
\ee
We claim that $\Phi(\cdot,t)$, and hence $\Phi_\#(\cdot,t)$, is 
regularly varying at $\infty$ with exponent 
$\hat\alpha=1/\alpha\in[\frac12,1)$. 

To prove the claim, we integrate by parts in \qref{psi0} to obtain
\[ \frac{\psi_0(q)}{q^2} = \frac{\sigma_0^2}{2}+ \int_0^\infty e^{-qs}
\left(\int_s^\infty \Ltail_0(r) \, dr \right) \, ds. \]
First assume $\sigma_0=0$ and $\Ltail_0$ is regularly varying at
zero with exponent $-\alpha$. 
Then $\psi_0$ is regularly varying at infinity with exponent $\alpha$. 
This follows from \cite[XIII.5.3]{Feller}, or 
may be proved directly.  If $\sigma_0 \neq 0$, we have
$\lim_{q \to \infty} \psi_0(q)/q^2=\sigma_0^2/2$. 
The Laplace exponent $\Phi(q,t)$ is determined via 
the functional relation \qref{E.main}. Since $\alpha >1$, the map 
$\Phi \mapsto g_0(\Phi):= \psi_0(t\Phi)+\Phi$ is regularly varying 
(in $\Phi)$ at $\infty$ with
exponent $\alpha$. Therefore, the inverse function $\Phi(q,t)$ is
regularly varying (in $q$) at $\infty$ with exponent $1/\alpha$.

Now let $g_\#(\cdot,t)$ be the inverse function to $\Phi_\#(\cdot,t)$.
Then by \cite[Lemma III.4.12]{B_book}
we infer that for every $\hat c>0$,
\be
\label{E.LIL8} 
 -\log \prob\left( \frac{k Y^t_{1/k}} {h_\#(k\log\log k,t)} \leq \hat{c} \right) \sim
(1-\hat\alpha)(\hat\alpha/\hat{c})^{{\hat\alpha}/(1-\hat\alpha)}
\log\log k, \quad k \to \infty,
\ee
where 
\[
h_\#(k,t) =  \frac{ k}{g_\#(k,t)}.
\]
By \qref{E.Phish} and regular variation we have 
$\Phi_\#(q,t)\sim \Phi(q/t,t)/\hat\alpha$ as $q\to\infty$,
and thus by \qref{E.main} we find that as $q\to\infty$,
\[
g_\#(q,t)\sim t g_0(\hat\alpha q)\sim t\psi_0(\hat\alpha tq)
\sim t^{1+\alpha}\alpha^{-\alpha} \psi_0(q).
\]
Substituting $\hat c=ct^{1+\alpha}\alpha^{-\alpha}$ into \qref{E.LIL8} 
yields Corollary~\ref{C.LIL}.

Karamata's Tauberian theorem and the monotone density theorem now imply that
$\Ltail_t(s)$ is regularly varying at zero with exponent
$-1/\alpha$. If $\sigma_0 \neq 0$, we find $\Phi(q,t) \sim (\sigma_0
t)^{-1} \sqrt{2q}$ as $q \to \infty$. Assertion (b) of the Lemma then follows from
the Tauberian theorem. 
\end{proof}

For the self-similar solutions, $\psi_0(q)=q^\alpha$ with $\alpha \in (1,2]$,
$\Ltail_0(s)=s^{-\alpha}/(\alpha\Gamma(-\alpha))$ for $\alpha\in(1,2)$, and we have
$\Ltail_t(s) \sim t^{-1}(ts)^{-1/\alpha}/\Gamma(1-1/\alpha)$ as $s \to 0$. 

\subsection{The multifractal spectrum}
The notion of a multifractal spectrum was introduced by Frisch and
Parisi to describe the intermittency of velocity fields in fully developed
turbulence~\cite{Frisch-Parisi}.  The multifractal spectrum $d(h)$ measures the 
dimension of the set $S_h$ where the velocity field has singularities of
order $h$. There are different mathematical formulations of
multifractality, corresponding to different notions of what one means
by singularities of order $h$. Here we follow the treatment by
Jaffard, which yields $d(h)$ rather easily~\cite{Jaffard} (the
notation has been changed slightly for consistency with this
article). 

We say a function
$v:\R_+ \to \R$, is $C^r(x_0)$ for a point $x_0 \in \R_+$ if there is
a polynomial $P_{x_0}$ of degree at most $[r]$ such that 
\[ |v(x)-P_{x_0}(x)| \leq C|x-x_0|^r, \]
in a neighborhood of $x_0$. The \Holder\/ exponent of $v$ at $x_0$ is
defined as 
\[ h_v(x_0)= \sup\{r\left| v \in C^r(x_0)\right. \}. \]
We define $S_h$ to be the set of points where $v$ is of \Holder\/
exponent $h$. The multifractal spectrum $d(h)$ is the Hausdorff
dimension of $S_h$. If $S_h$ is empty, the convention is
$d(h)=-\infty$. As an example, let us compute the multifractal
spectrum when the initial data is of bounded variation. Then $M_0< \infty$ 
in~\qref{E.BV} and there is a finite number of 
shocks $s^t_y$ in a finite interval $[0,x]$ with probability
$1$. Suppose $x_0$ is not a shock location for $v(\cdot,t)$. Then~\qref{E.BV}
shows $v$ is analytic near $x_0$ and  $h_v(x_0) = \infty$.  If $x_0$ is a shock
location, then $v \in C^{-\eps}(x_0)$ for every $\eps >0$, so that
$h_v(x_0)=0$. Thus, we have 
simply $d(0)=0$ and $d(h)= -\infty$ for every $h \neq 0$. 

The multifractal spectrum is more interesting for initial data of
unbounded variation, that is, when \qref{E.sigma1} holds. In this case, the
jumps in $v$ are dense. Following Jaffard~\cite{Jaffard}, 
the multifractal spectrum is computed as follows. We define
\be
\label{E.multfrac}
C_j(t) = \int_{2^{-j-1}}^{2^{-j}} \Lambda_t(ds), \quad
\spec_t = \max \left(0, \limsup_{j \to \infty} \frac{\log C_j(t)}{j
  \log 2} \right). 
\ee
For any $t >0$, $v(\cdot, t)$ has no Brownian component. It then
follows from~\cite[Thm. 1]{Jaffard} that 
\be
\label{E.multfrac2}
 d_t(h) = \left\{ \begin{array}{ll} \spec_th, &\quad  h \in [0,
    1/\spec_t], \\ -\infty, &\quad \mathrm{else}. \end{array} \right.
\ee
Experiments suggest that the multifractal spectrum for fully developed
three-dimensional turbulence is a concave
curve~\cite[Fig. 2]{Meneveau-Sreenivasan}. This is in clear contrast
with \qref{E.multfrac2}.

For example, let us compute the multifractal spectrum for the
self-similar process $V^\alpha$ of index 
$\alpha \in (1,2]$. Since $\Lambda^\alpha_t(s)$ is a
scaled copy of $\Lambda^\alpha_1(s)$, $d_t(h)$ is independent of $t$. 
We use \qref{E.ndensity} to obtain the asymptotics as $s \to
0$: 
\[ \Lambda^\alpha_1(ds) = f_\alpha(s) \, ds \sim \frac{\sin \pi
  \beta}{\pi} s^{\beta-2} 
\Gamma(2-\beta) \, ds ,  \quad \beta = \frac{\alpha-1}{\alpha}. \]
We then have $\spec_t= \alpha^{-1}, t >0$ and 
\be
\label{E.multfrac3}
d(h) = \left\{ \begin{array}{ll} h/\alpha, &\quad  h \in [0, \alpha],
  \\ -\infty, &\quad \mathrm{else}. \end{array} \right. 
\ee
In particular, \qref{E.multfrac3} implies that $d(\alpha)=1$, that is
$v(x,t)$ is $C^{\alpha}(x)$ for a.e $x \in \R_+$. For this set a finer
characterization of the local variation of $v(\cdot,t)$ may be
obtained by using the Fristedt-Pruitt  law of the iterated
logarithm (see~\cite[Cor.~1]{B_burgers}). However, the
multifractal spectrum, also describes sets
$S_h$, $0<h< \alpha$, that are not covered by the Fristedt-Pruitt law.

\section{Acknowledgement}
This material is based upon work supported by the National Science Foundation
under grants DMS 03-05985 and DMS 04-05343. G.M. thanks the University
of Crete for hospitality during part of this work. 

\bibliographystyle{siam}
\bibliography{mp4}

\begin{thebibliography}{10}

\bibitem{Abramowitz}
{\sc M.~Abramowitz and I.~A. Stegun}, {\em Handbook of mathematical functions
  with formulas, graphs, and mathematical tables}, vol.~55 of National Bureau
  of Standards Applied Mathematics Series, For sale by the Superintendent of
  Documents, U.S. Government Printing Office, Washington, D.C., 1964.

\bibitem{Aldous}
{\sc D.~J. Aldous}, {\em Deterministic and stochastic models for coalescence
  (aggregation and coagulation): a review of the mean-field theory for
  probabilists}, Bernoulli, 5 (1999), pp.~3--48.

\bibitem{AE}
{\sc M.~Avallaneda and W.~E}, {\em Statistical propeties of shocks in {B}urgers
  turbulence}, Comm. Math. Phys., 172 (1995), pp.~13--38.

\bibitem{B_book}
{\sc J.~Bertoin}, {\em L\'evy processes}, vol.~121 of Cambridge Tracts in
  Mathematics, Cambridge University Press, Cambridge, 1996.

\bibitem{B_burgers}
\leavevmode\vrule height 2pt depth -1.6pt width 23pt, {\em The inviscid
  {B}urgers equation with {B}rownian initial velocity}, Comm. Math. Phys., 193
  (1998), pp.~397--406.

\bibitem{B_cluster}
\leavevmode\vrule height 2pt depth -1.6pt width 23pt, {\em Clustering
  statistics for sticky particles with {B}rownian initial velocity}, J. Math.
  Pures Appl. (9), 79 (2000), pp.~173--194.

\bibitem{B_eternal}
\leavevmode\vrule height 2pt depth -1.6pt width 23pt, {\em Eternal solutions to
  {S}moluchowski's coagulation equation with additive kernel and their
  probabilistic interpretations}, Ann. Appl. Probab., 12 (2002), pp.~547--564.

\bibitem{B_icm}
\leavevmode\vrule height 2pt depth -1.6pt width 23pt, {\em Some aspects of
  additive coalescents}, in Proceedings of the International Congress of
  Mathematicians, Beijing 2002, vol.~III, Higher Ed. Press, 2002, pp.~15--23.

\bibitem{Bingham2}
{\sc N.~H. Bingham}, {\em Maxima of sums of random variables and suprema of
  stable processes}, Z. Wahrscheinlichkeitstheorie und Verw. Gebiete, 26
  (1973), pp.~273--296.

\bibitem{Burgers}
{\sc J.~M. Burgers}, {\em The nonlinear diffusion equation}, Dordrecht: Reidel,
  1974.

\bibitem{Duchon0}
{\sc L.~Carraro and J.~Duchon}, {\em Solutions statistiques intrins\`eques de
  l'\'equation de {B}urgers et processus de {L}\'evy}, C. R. Acad. Sci. Paris
  S\'er. I Math., 319 (1994), pp.~855--858.

\bibitem{Duchon1}
\leavevmode\vrule height 2pt depth -1.6pt width 23pt, {\em \'{E}quation de
  {B}urgers avec conditions initiales \`a accroissements ind\'ependants et
  homog\`enes}, Ann. Inst. H. Poincar\'e Anal. Non Lin\'eaire, 15 (1998),
  pp.~431--458.

\bibitem{Duchon2}
{\sc M.-L. Chabanol and J.~Duchon}, {\em Markovian solutions of inviscid
  {B}urgers equation}, J. Statist. Phys., 114 (2004), pp.~525--534.

\bibitem{Chassaing}
{\sc P.~Chassaing and G.~Louchard}, {\em Phase transition for parking blocks,
  {B}rownian excursion and coalescence,}, Random Structures and Algorithms, 21
  (2002), pp.~76--119.

\bibitem{Cole}
{\sc J.~D. Cole}, {\em On a quasi-linear parabolic equation occurring in
  aerodynamics}, Quart. Appl. Math., 9 (1951), pp.~225--236.

\bibitem{Drake}
{\sc R.~L. Drake}, {\em A general mathematical survey of the coagulation
  equation}, in Topics in Current Aerosol Research, G.~M. Hidy and J.~R. Brock,
  eds., no.~2 in International reviews in Aerosol Physics and Chemistry,
  Pergammon, 1972, pp.~201--376.

\bibitem{ES}
{\sc W.~E and Y.~G. Sina\u{\i}}, {\em New results in mathematical and
  statistical hydrodynamics}, Uspekhi Mat. Nauk, 55 (2000), pp.~25--58.

\bibitem{Feller}
{\sc W.~Feller}, {\em An introduction to probability theory and its
  applications. {V}ol. {II}.}, Second edition, John Wiley \& Sons Inc., New
  York, 1971.

\bibitem{Frachebourg}
{\sc L.~Frachebourg and P.~A. Martin}, {\em Exact statistical properties of the
  {B}urgers equation}, J. Fluid Mech., 417 (2000), pp.~323--349.

\bibitem{Frisch-Parisi}
{\sc U.~Frisch and G.~Parisi}, {\em On the singularity structure of fully
  developed turbulence}, in Turbulence and predictability in geophysics,
  M.~Ghil, R.~Benzi, and R.Parisi, eds., North-Holland, 1985, pp.~84--87.

\bibitem{Woc2}
{\sc T.~Funaki, D.~Surgailis, and W.~A. Woyczy{\'n}ski}, {\em Gibbs-{C}ox
  random fields and {B}urgers turbulence}, Ann. Appl. Probab., 5 (1995),
  pp.~461--492.

\bibitem{Giraud}
{\sc C.~Giraud}, {\em Genealogy of shocks in {B}urgers turbulence with white
  noise initial velocity}, Comm. Math. Phys., 223 (2001), pp.~67--86.

\bibitem{Golovin}
{\sc A.~M. Golovin}, {\em The solution of the coagulating equation for cloud
  droplets in a rising air current}, Izv. Geophys. Ser.,  (1963), pp.~482--487.

\bibitem{Groen}
{\sc P.~Groeneboom}, {\em Brownian motion with a parabolic drift and {A}iry
  functions}, Probab. Theory Related Fields, 81 (1989), pp.~79--109.

\bibitem{Gurbatov2}
{\sc S.~N. Gurbatov, A.~N. Malakhov, and A.~I. Saichev}, {\em Nonlinear random
  waves and turbulence in nondispersive media: waves, rays, particles},
  Manchester University Press, Manchester, 1991.

\bibitem{Gurbatov}
{\sc S.~N. Gurbatov, S.~I. Simdyankin, E.~Aurell, U.~Frisch, and G.~T{\'o}th},
  {\em On the decay of {B}urgers turbulence}, J. Fluid Mech., 344 (1997),
  pp.~339--374.

\bibitem{Hopf}
{\sc E.~Hopf}, {\em The partial differential equation {$u\sb t+uu\sb x=\mu\sb
  {xx}$}}, Comm. Pure Appl. Math., 3 (1950), pp.~201--230.

\bibitem{Jacod}
{\sc J.~Jacod and A.~N. Shiryaev}, {\em Limit theorems for stochastic
  processes}, vol.~288 of Grundlehren der Mathematischen Wissenschaften,
  Springer-Verlag, Berlin, second~ed., 2003.

\bibitem{Jaffard}
{\sc S.~Jaffard}, {\em The multifractal nature of {L}\'evy processes}, Probab.
  Theory Related Fields, 114 (1999), pp.~207--227.

\bibitem{Kolm}
{\sc A.~N. Kolmogorov}, {\em Dissipation of energy in the locally isotropic
  turbulence}, Proc. Roy. Soc. London Ser. A, 434 (1991), pp.~15--17.
\newblock Translated from the Russian by V. Levin, Turbulence and stochastic
  processes: Kolmogorov's ideas 50 years on.

\bibitem{Meneveau-Sreenivasan}
{\sc C.~Meneveau and K.~R. Sreenivasan}, {\em Simple multifractal cascade model
  for fully developed turbulence}, Phys. Rev. Lett., 59 (1987), pp.~1424--1427.

\bibitem{MP1}
{\sc G.~Menon and R.~Pego}, {\em Approach to self-similarity in
  {S}moluchowski's coagulation equations}, Comm. Pure Appl. Math., 57 (2004),
  pp.~1197--1232.

\bibitem{Resnick}
{\sc S.~Resnick}, {\em Extreme values, regular variation and point processes},
  Springer-Verlag, New York, 1987.

\bibitem{She}
{\sc Z.-S. She, E.~Aurell, and U.~Frisch}, {\em The inviscid {B}urgers equation
  with initial data of {B}rownian type}, Comm. Math. Phys., 148 (1992),
  pp.~623--641.

\bibitem{Sinai}
{\sc Y.~G. Sina\u{\i}}, {\em Statistics of shocks in solutions of inviscid
  {B}urgers equation}, Comm. Math. Phys., 148 (1992), pp.~601--621.

\bibitem{Woc}
{\sc W.~A. Woyczy{\'n}ski}, {\em Burgers-{KPZ} turbulence}, vol.~1700 of
  Lecture Notes in Mathematics, Springer-Verlag, Berlin, 1998.
\newblock G\"ottingen lectures.

\bibitem{Ziff}
{\sc R.~M. Ziff}, {\em Kinetics of polymerization}, J. Statist. Phys., 23
  (1980), pp.~241--263.

\end{thebibliography}
\end{document}